\documentclass[pra, twocolumn]{revtex4}
\usepackage[ansinew]{inputenc}
\usepackage{graphicx}
\usepackage{amsmath}
\usepackage{amsthm}
\usepackage{bm}
\usepackage{layout}
\usepackage{float}
\usepackage{amsfonts}
\usepackage{amssymb}
\usepackage{txfonts}%
\usepackage{hyperref}
\newcommand\id{\leavevmode\hbox{\small1\kern-3.3pt\normalsize1}}

\newcommand{\bra}{\langle}
\newcommand{\ket}{\rangle}

\newcommand{\tr}{\mbox{Tr}}

\begin{document}

\title{Quantum correlations with no causal order\footnote{The published version of this paper can be found in \textit{Nature Communications} \textbf{3}, 1092 (2012), doi:10.1038/ncomms2076, at \href{http://www.nature.com/ncomms/journal/v3/n10/full/ncomms2076.html}{http://www.nature.com/ncomms/journal/v3/n10/full/ncomms2076.html}.}}


\author{Ognyan Oreshkov$^{1,2}$, Fabio Costa$^1$, {\v C}aslav Brukner$^{1,3}$}

\affiliation{
$^1$Faculty of Physics, University of Vienna, Boltzmanngasse 5,
A-1090 Vienna, Austria.\\
$^2$QuIC, Ecole Polytechnique, CP 165, Universit\'{e} Libre de Bruxelles, 1050 Brussels, Belgium.\\
$^3$Institute of Quantum Optics and Quantum Information, Austrian Academy of Sciences, Boltzmanngasse 3,
A-1090 Vienna, Austria.}

\begin{abstract}
The idea that events obey a definite causal order is deeply rooted in our understanding of the world and at the basis of the very notion of time. But where does causal order come from, and is it a necessary property of nature? We address these questions from the standpoint of quantum mechanics in a new framework for multipartite correlations which does not assume a pre-defined global causal structure but only the validity of quantum mechanics locally. All known situations that respect causal order, including space-like and time-like separated experiments, are captured by this framework in a unified way. Surprisingly, we find correlations that cannot be understood in terms of definite causal order. These correlations violate a `causal inequality' that is satisfied by all space-like and time-like correlations. We further show that in a classical limit causal order always arises, which suggests that space-time may emerge from a more fundamental structure in a quantum-to-classical transition.

\end{abstract}

\maketitle

\section*{\normalsize{Introduction}}

One of the striking features of quantum mechanics is that it challenges the view that physical properties are well defined prior to and independent of their measurement. This motivates an operational approach to the theory, where primitive laboratory procedures, such as measurements and preparations, are basic ingredients. Although significant progress has recently been made in this direction \cite{Hardy, Zeilinger, CBH, goyal, DB, Mas, Fivel, Chiribella2}, most approaches still retain a notion of space-time as a pre-existing `stage' in which events take place. Even the most abstract constructions, in which no explicit reference to space-time is made, do assume a definite order of events: if a signal is sent from an event $A$ to an event $B$ in the run of an experiment, no signal can be sent in the opposite direction in that same run.
But are space, time, and causal order truly fundamental ingredients of nature? Is it possible that, in some circumstances, even causal relations would be `uncertain', similarly to the way other physical properties of quantum systems are \cite{hardyqg}?

Here we show that quantum mechanics allows for such a possibility. We develop a framework that describes all correlations that can be observed by two experimenters under the assumption that in their local laboratories physics is described by the standard quantum formalism, but without assuming that the laboratories are embedded in any definite causal structure. These include non-signalling correlations arising from measurements on a bipartite state, as well as signalling ones, which can arise when a system is sent from one laboratory to another through a quantum channel. We find that, surprisingly, more general correlations are possible, which are not included in the standard quantum formalism. These correlations are incompatible with any underlying causal structure: they allow performing a task---the violation of a `causal inequality'---which is impossible if events take place in a causal sequence. This is directly analogous to the famous violation of local realism: quantum systems allow performing a task---the violation of Bell's inequality \cite{Bell}---which is impossible if the measured quantities have pre-defined local values. The inequality considered here, unlike Bell's, concerns signalling correlations: it is based on a task that involves communication between two parties. Nevertheless, it cannot be violated if this communication takes place in a causal space-time. Previous works about relativistic causality in quantum mechanics focused on non-signalling correlations between space-like separated experiments or on a finite speed of signalling \cite{PR, BGNP, BHK, werner, infocaus, macrolocality, wolf, Barnum, Acin}. In the present work we go beyond such approaches since we do not assume the existence of a space-time (or more generally of a definite causal structure) on which the evolution of quantum systems and the constraints given by relativity are defined. One of the motivations for our approach comes from the problem of time in attempts to merge quantum theory and general relativity into a more fundamental theory \cite{dewitt, perestime, wooters, isham, rovelli, gambini}.

\section*{\normalsize{Results}}

\subsection*{Causal inequality}

The general setting that we consider involves a number of experimenters---Alice, Bob and others---who reside in separate laboratories. At a given run of the experiment, each of them receives a physical system (for instance, a spin-$\frac{1}{2}$ particle) and performs operations on it (e.g. measurements or rotations of the spin), after which she/he sends the system out of the laboratory. We assume that during the operations of each experimenter, the respective laboratory is isolated from the rest of the world---it is only opened for the system to come in and to go out, but between these two events it is kept closed. It is easy to see that, under this assumption, causal order puts a restriction on the way in which the parties can communicate during a given run. For instance, imagine that Alice can send a signal to Bob. [Formally, sending a signal (or signalling) is the existence of statistical correlations between a random variable that can be chosen by the sender and another one observed by the receiver.] Since Bob can only receive a signal through the system entering his laboratory, this means that Alice must act on her system before that. But this implies that Bob cannot send a signal to Alice since each party receives a system only once. Therefore, bidirectional signalling is forbidden.

Consider, in particular, the following communication task to be performed by two parties, Alice and Bob. After a given party receives the system in her/his laboratory, she/he will have to toss a coin (or use any other means) to obtain a random bit. Denote the bits generated by Alice and Bob in this way by $a$ and $b$, respectively. In addition, Bob will have to generate another random bit $b'$, whose value, $0$ or $1$, will specify their goal: if $b'=0$, Bob will have to communicate the bit $b$ to Alice, while if $b'=1$, he will have to guess the bit $a$. Without loss of generality, we will assume that the parties always produce a guess, denoted by $x$ and $y$ for Alice and Bob respectively, for the bit of the other (although the guess may not count depending on the value of $b'$). Their goal is to maximize the probability of success
\begin{gather}
p_{succ}:=\frac{1}{2}\left[ P(x=b|b'=0)+P(y=a|b'=1)\right].
\end{gather}

\begin{figure}
\begin{center}
\includegraphics[width=8.5cm]{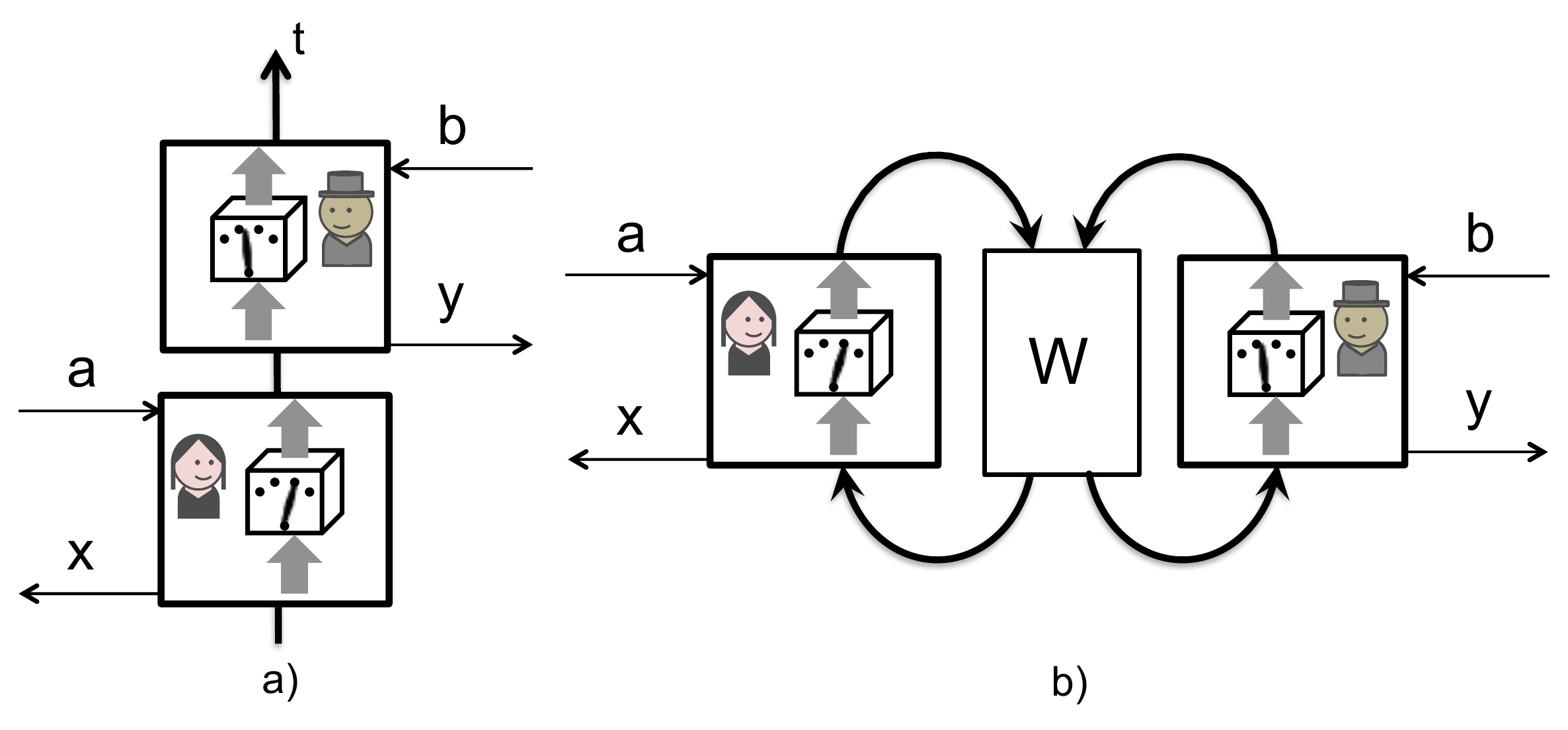}
\end{center}
\vspace{-0.5 cm}
\caption{\textbf{Strategy for accomplishing communication task by using processes with definite and indefinite causal order.} (a) There exists a global background time according to which Alice's actions are strictly before Bob's. She sends her input $a$ to Bob, who can read it out at some later time and give his estimate $y=a$. However, Bob cannot send his bit $b$ to Alice as the system passes through her laboratory at some earlier time. Consequently, she can only make a random guess of Bob's bit. This results in a probability of success of $3/4$. (b) If the assumption of a definite order is dropped, it is possible to devise a resource (i.e. a process matrix $W$) and a strategy that enables a probability of success $\frac{2+\sqrt{2}}{4}>3/4$ (see text).} \label{task}
\end{figure}

If all events obey causal order, no strategy can allow Alice and Bob to exceed the bound
\begin{gather}
p_{succ}\leq 3/4.\label{bound}
\end{gather}
Indeed, as argued above, in any particular order of events, there can be at most unidirectional signalling between the parties, which means that at least one of the following must be true: Alice cannot signal to Bob, or Bob cannot signal to Alice. Consider, for example, a case where Bob cannot signal to Alice. Then, if $b'=1$, they could in principle achieve up to $P(y=a|b'=1)=1$ (for instance, if Alice operates on her system before Bob, she could encode information about the bit $a$ in the system and send it to him). However, if $b'=0$, the best guess that Alice can make is a random one, resulting in $P(x=b|b'=0)=1/2$ (see Fig.~\ref{task}a). Hence, the overall probability of success in this case will satisfy $p_{succ}\leq 3/4$. The same holds if Alice cannot signal to Bob. It is easy to see that no probabilistic strategy can increase the probability of success.

\renewcommand{\figurename}{{Fig.}}

Formally, the assumptions behind the causal inequality \eqref{bound} can be summarized as follows:

\emph{Causal structure} (\emph{CS})---The main events in the task (a system entering Alice's/Bob's laboratory, the parties obtaining the bits $a$, $b$, and $b'$, and producing the guesses $x$ and $y$) are localized in a causal structure. [A causal structure (such as space-time) is a set of event locations equipped with a partial order $\preceq$ that defines the possible directions of signalling. If $A\preceq B$, we say that $A$ is in the \textit{causal past} of $B$ (or $B$ is in the \textit{causal future} of $A$). In this case, signalling from $A$ to $B$ is possible, but not from $B$ to $A$. For more details on causal structures, see Appendix.]

\emph{Free choice} (\emph{FC})---Each of the bits $a$, $b$, and $b'$ can only be correlated with events in its causal future (this concerns only events relevant to the task). We assume also that each of them takes values $0$ or $1$ with probability $1/2$.

\emph{Closed laboratories} (\emph{CL})---Alice's guess $x$ can be correlated with Bob's bit $b$ only if the latter is generated in the causal past of the system entering Alice's laboratory. Analogously, $y$ can be correlated with $a$ only if $a$ is generated in the causal past of the system entering Bob's laboratory.

In the Appendix, we present a formal derivation of the inequality from these assumptions.

Interestingly, we will see that if the local laboratories are described by quantum mechanics, but no assumption about a global causal structure is made  (Fig.~\ref{task}b), it is in principle possible to violate the causal inequality in physical situations in which one would have all the reasons to believe that the bits are chosen freely and the laboratories are closed. This would imply that \emph{CS} does not hold.

\subsection*{Framework for local quantum mechanics}

\begin{figure}
\begin{center}
\includegraphics[width=8cm]{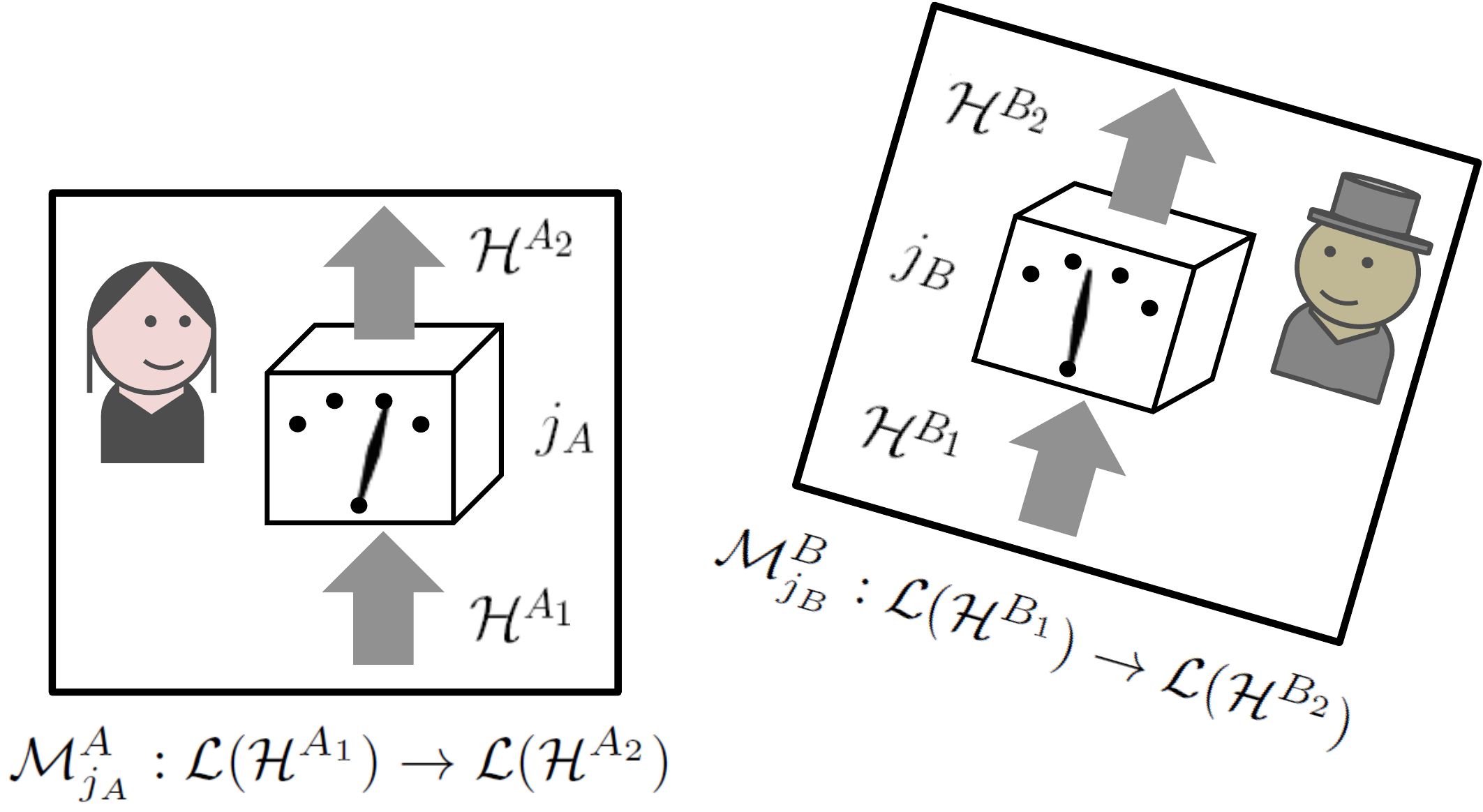}
\end{center}
\vspace{-0.5 cm}
\caption{\textbf{Local quantum experiments with no assumption of a pre-existing background time or global causal structure.} While the global causal order of events in the two laboratories is not fixed in advance and in general not even definite (here illustrated by the `shifted' relative orientation of the two laboratories), the two agents, Alice and Bob, are each certain about the causal order of events in their respective laboratories.} \label{stage}
\end{figure}

The most studied, almost epitomical, quantum correlations are the non-signalling ones, such as those obtained when Alice and Bob perform measurements on two entangled systems. Signalling quantum correlations exist as well, such as those arising when Alice operates on a system which is subsequently sent through a quantum channel to Bob who operates on it after that. The usual quantum formalism does not consider more general possibilities, since it does assume a global causal structure. Here we want to drop the latter assumption while retaining the validity of quantum mechanics locally. For this purpose, we consider a multipartite setting of the type outlined earlier, where each party performs an operation on a system passing once through her/his laboratory, but we make no assumption about the spatio-temporal location of these experiments, not even that there exists a space-time or any causal structure in which they could be positioned (see Fig.~\ref{stage}). Our framework is thus based on the following central premise:

\begin{quote}
\textit{Local quantum mechanics}---The local operations of each party are described by quantum mechanics.
\end{quote}

More specifically, we assume that one party, say Alice, can perform all the operations she could perform in a closed laboratory, as described in the standard space-time formulation of quantum mechanics. These are defined as the set of \textit{quantum instruments} \cite{instrument} with an input Hilbert space ${\cal H}^{A_1}$ (the system coming in) and an output Hilbert space ${\cal H}^{A_2}$ (the system going out). (The set of allowed quantum operations can be used as a definition of `closed quantum laboratory' with no reference to a global causal structure.) A quantum instrument can most generally be realized by applying a joint unitary transformation on the input system plus an ancilla, followed by a projective measurement on part of the resulting joint system, which leaves the other part as an output. (From the point of view of each party, the input/output systems most generally correspond to two subsystems of the Hilbert space associated with the local laboratory, each considered at a different instant---the time of entrance and the time of exit, respectively---where the subsystems and the respective instants are independent of the choice of operation that connects them.) When Alice uses a given instrument, she registers one out of a set of possible outcomes, labeled by $j=1,\dots,n$. Each outcome induces a specific transformation from the input to the output, which corresponds to a completely positive (CP) trace-nonincreasing map \cite{nielsen_chuang} ${\cal M}_j^A:{\cal L}({\cal H}^{A_1})\rightarrow {\cal L}({\cal H}^{A_2})$, where ${\cal L}({\cal H}^X)$, $X=A_1, A_2$, is the space of matrices over a Hilbert space ${\cal H}^X$ of dimension $d_X$. The action of each ${\cal M}_j^A$ on any matrix $\sigma\in {\cal L}({\cal H}^{A_1})$ can be written as \cite{nielsen_chuang} $\mathcal{M}_j^A(\sigma)=\sum_{k=1}^{m}E_{jk}\sigma E_{jk}^{\dagger}$, $m = d_{A_1}d_{A_2}$, where the matrices $E_{jk}:{\cal H}^{A_1}\rightarrow {\cal H}^{A_2}$ satisfy $\sum_{k=1}^{m} E^{\dagger}_{jk}E_{jk}\leq \id^{A_1}$, $\forall j$.
If the operation is performed on a quantum state described by a density matrix $\rho$, $\mathcal{M}_j^A(\rho)$ describes the updated state after the outcome $j$ up to normalization, while the probability to observe this outcome is given by $P\left(\mathcal{M}_j^A \right)= \tr\left[\mathcal{M}_j^A(\rho)\right]$. The set of CP maps $\left\{{\cal M}_j^A\right\}_{j=1}^n$ corresponding to all the possible outcomes of a quantum instrument has the property that $\sum_{j=1}^n{\cal M}_j^A$ is CP and trace-preserving (CPTP), or equivalently $\sum_{j=1}^{n}\sum_{k=1}^{m}E^{\dagger}_{jk}E_{jk} = \id^{A_1}$, which reflects the fact that the probability to observe any of the possible outcomes is unity. A CPTP map itself corresponds to an instrument with a single outcome which occurs with certainty.

In the case of more than one party, the set of local outcomes corresponds to a set of CP maps ${\cal M}_i^A, {\cal M}_j^B, \cdots$. A complete list of probabilities $P\left(\mathcal{M}^A_i,\mathcal{M}^B_j,\cdots\right)$ for all possible local outcomes will be called \textit{process}. (It is implicitly assumed that the joint probabilities are {noncontextual}, namely that they are independent of any variable concerning the concrete implementation of the local CP maps. For example, the probability for a pair of maps $\mathcal{M}^A_i$, $\mathcal{M}^B_j$ to be realized should not depend on the particular set $\{\mathcal{M}^A_1,...,\mathcal{M}^A_i,...,\mathcal{M}^A_{n}\}$ of possible CP maps associated with Alice's operation.) A process can be seen as an extension of the notion of state as a list of probabilities for detection results \cite{Hardy} described by a positive operator-valued measure (POVM), which takes into account the transformation of the system after the measurement and can thus capture more general scenarios than just detection. Here we will consider explicitly only the case of two parties (the generalization to arbitrarily many parties is straightforward). We want to characterize the most general probability distributions for a pair of outcomes $i$, $j$, corresponding to CP maps $\mathcal{M}^A_{i}$, $\mathcal{M}^B_{j}$, to be observed, that is, to characterize all bipartite processes.

In quantum mechanics, operations obey a specific algebraic structure that reflects the operational relations between laboratory procedures \cite{Hardy}. For example, a probabilistic mixture of operations is expressed as a linear convex combination of CP maps. It can be shown (see Appendix) that the only probabilities $P\left(\mathcal{M}^A_i,\mathcal{M}^B_j\right)$ consistent with the algebraic structure of local quantum operations are bilinear functions of the CP maps $\mathcal{M}^A_i$ and $\mathcal{M}^B_j$. Thus the study of the most general bipartite quantum correlations reduces to the study of bilinear functions of CP maps.

It is convenient to represent CP maps by positive semi-definite matrices via the  Choi-Jamio{\l}kowsky (CJ) isomorphism \cite{jam, choi}. The CJ matrix $M^{A_1A_2}_i\in{\cal L}({\cal H}^{A_1}\otimes{\cal H}^{A_2})$ corresponding to a linear map ${\cal M}_i:{\cal L}({\cal H}^{A_1})\rightarrow {\cal L}({\cal H}^{A_2})$ is defined as $M^{A_1A_2}_i:=\left[{\cal I}\otimes{\cal M}_i\left(|\phi^+\ket\bra \phi^+|\right)\right]^{\mathrm T}$, where $|\phi^+\ket=\sum_{j=1}^{d_{A_1}}|jj\ket \in {\cal H}^{A_1}\otimes{\cal H}^{A_1}$ is a (not normalized) maximally entangled state, the set of states $\left\{|j\ket\right\}_{j=1}^{d_{A_1}}$ is an orthonormal basis of ${\cal H}^{A_1}$, ${\cal I}$ is the identity map, and ${\mathrm T}$ denotes matrix transposition (the transposition, absent in the original definition, is introduced for later convenience). Using this correspondence, the probability for two measurement outcomes can be expressed as a bilinear function of the corresponding CJ operators as follows:
\begin{equation}
	\label{representation}
	P\left(\mathcal{M}^A_i, \mathcal{M}^B_j\right) = \tr \left[W^{A_1A_2B_1B_2}\left(M^{A_1A_2}_i\otimes M_j^{B_1B_2}\right)\right],
\end{equation}
where $W^{A_1A_2B_1B_2}$ is a matrix in ${\cal L}({\cal H}^{A_1}\otimes{\cal H}^{A_2}\otimes{\cal H}^{B_1}\otimes{\cal H}^{B_2})$.

The matrix $W$ should be such that probabilities are non-negative for any pair of CP maps $\mathcal{M}^A_i$, $\mathcal{M}^B_j$. We require that this be true also for measurements in which the system interacts with any system in the local laboratory, including systems entangled with the other laboratory. This implies that $W^{A_1A_2B_1B_2}$ must be positive semidefinite (see Appendix). Furthermore, the probability for any pair of CPTP maps $\mathcal{M}^A$, $\mathcal{M}^B$ to be realized must be unity (they correspond to instruments with a single outcome). Since a map $\mathcal{M}^A$ is CPTP if and only if its CJ operator satisfies $M^{A_1A_2}\geq 0$ and $\tr_{A_2}M^{A_1A_2}=\id^{A_1}$ (similarly for $\mathcal{M}^B$), we conclude that all bipartite probabilities compatible with local quantum mechanics are generated by matrices $W$ that satisfy
\begin{gather}
 W^{A_1A_2B_1B_2} \geq 0 \hspace{0.2cm} [\mbox{{\em non-negative probabilities}}],\label{W1} \\
 \tr \left[W^{A_1A_2B_1B_2}\left(M^{A_1A_2}\otimes M^{B_1B_2}\right)\right] = 1, \notag
\\
\forall M^{A_1A_2},M^{B_1B_2}\geq 0,\hspace{0.1cm} \tr_{A_2}M^{A_1A_2}=\id^{A_1}, \tr_{B_2}M^{B_1B_2}=\id^{B_1}\label{W2}\\
 [\mbox{{\em probabilities sum up to 1}}].\notag
\end{gather}
We will refer to a matrix $W^{A_1A_2B_1B_2}$ that satisfies these conditions as a \textit{process matrix}.
Conditions equivalent to Eqs.~\eqref{W1} and \eqref{W2} were first derived as part of the definition of a `quantum comb' \cite{networks}, an object that formalizes quantum networks. Combs, however, are subject to additional conditions fixing a definite causal order, which are not assumed here.

A process matrix can be understood as a generalization of a density matrix and Eq.~\eqref{representation} can be seen as a generalization of Born's rule. In fact, when the output systems $A_2$, $B_2$ are taken to be one-dimensional (i.e. each party performs a measurement after which the system is discarded), the expression above reduces to $P\left(\mathcal{M}^A_i, \mathcal{M}^B_j\right) = \tr \left[W^{A_1B_1}\left(M^{A_1}_i\otimes M_j^{B_1}\right)\right]$, where now $M^A_i, M^B_j$ are elements of local POVMs and $W^{A_1B_1}$ is a quantum state. This implies that a quantum state $\rho^{A_1B_1}$ shared by Alice and Bob is generally represented by the process matrix $W^{A_1A_2B_1B_2} =\rho^{A_1B_1}\otimes\id^{A_2B_2}$. Signalling correlations can also be expressed in terms of process matrices. For instance, the situation where Bob is given a state $\rho^{B_1}$ and his output is sent to Alice through a quantum channel ${\cal C}$, which gives $P(\mathcal{M}^A_i,\mathcal{M}^B_j) =\tr\left[\mathcal{M}^A_i\circ\mathcal{C}\circ\mathcal{M}^B_j\left(\rho^{B_1}\right)\right]$, is described by $W^{A_1A_2B_1B_2}= \id^{A_2}\otimes (C^{B_2A_1})^{T}\otimes \rho^{B_1}$, where $C^{B_2A_1}$ is the CJ matrix of the channel ${\cal C}$ from $B_2$ to $A_1$.

The most general bipartite situation typically encountered in quantum mechanics (i.e. one that can be expressed in terms of a quantum circuit) is a quantum channel with memory, where, say, Bob operates on one part of an entangled state and his output plus the other part is transferred to Alice through a channel. This is described by a process matrix of the form $\id^{A_2}\otimes W^{A_1B_1B_2}$. Conversely, all process matrices of this form represent channels with memory \cite{networks}. This is the most general situation in which signalling from Alice to Bob is not possible, a relation that we will denote by $A \npreceq B$ in accord with the causal notation introduced earlier. Process matrices of this kind will be denoted by $W^{A\npreceq B}$ (note that for non-signalling processes, both $A\npreceq B$ and $B\npreceq A$ are true). As argued earlier, if all events are localized in a causal structure and Alice and Bob perform their experiments inside closed laboratories, at most unidirectional signalling between the laboratories is allowed. In a definite causal structure, it may still be the case that the location of each event, and thus the causal relation between events, is not known with certainty. A situation where $B\npreceq A$ with probability $0\leq q \leq 1$ and $A\npreceq B$ with probability $1-q$ is represented by a process matrix of the form
\begin{equation} \label{CS}
W^{A_1A_2B_1B_2} = q W^{B \npreceq A} + (1-q) W^{A\npreceq B}.
\end{equation}
We will call processes of this kind \textit{causally separable} (note that the decomposition \eqref{CS} need not be unique since non-signalling processes can be included either in $W^{B \npreceq A}$ or in $W^{A\npreceq B}$). They represent the most general bipartite quantum processes for which the local experiments are performed in closed laboratories embedded in a definite causal structure. In particular, they generate the most general quantum correlations between measurements that take place at definite (though possibly unknown) instants of time. Clearly, according to the argument presented earlier, causally separable processes cannot be used by Alice an Bob to violate the causal inequality \eqref{bound}.

\begin{figure}
\begin{center}
\includegraphics[width=8.7cm]{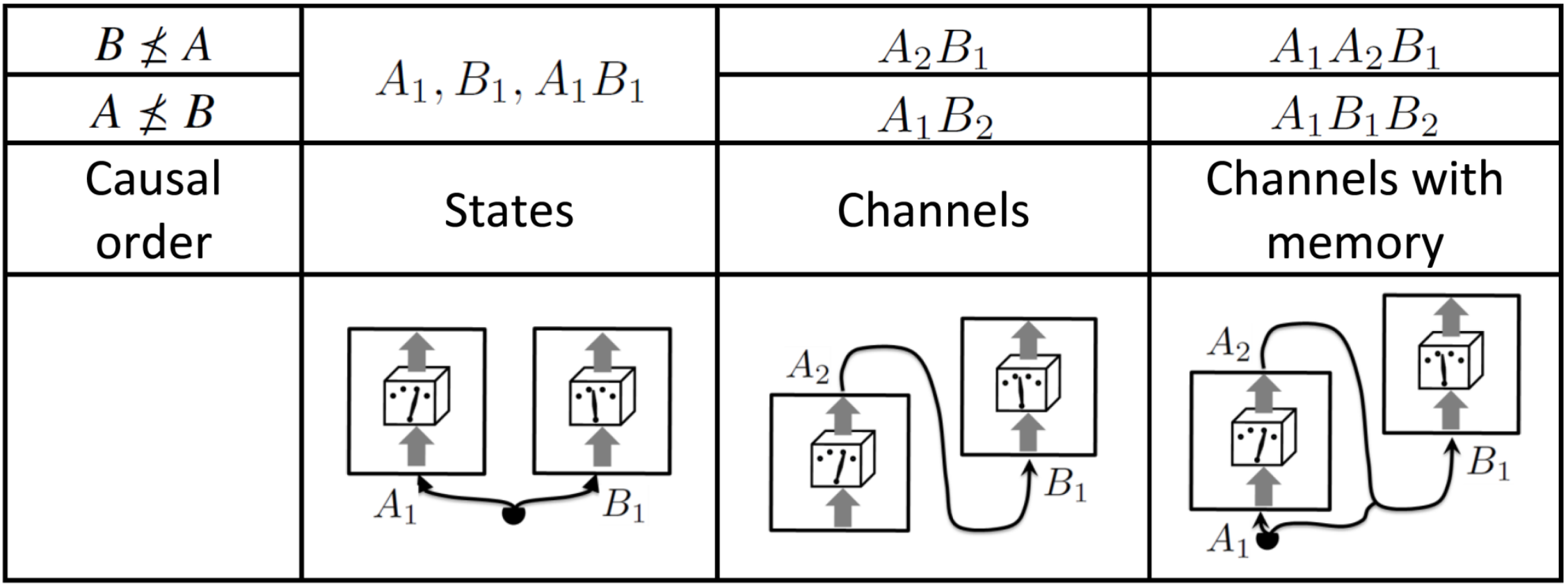}
\end{center}
\vspace{-0.5 cm}
\caption{\textbf{Terms appearing in a process matrix.} A matrix satisfying condition \eqref{W1} can be expanded as $
W^{A_1A_2B_1B_2}= \sum_{\mu \nu \lambda \gamma} w_{\mu \nu \lambda \gamma} \sigma^{A_1}_{\mu}\otimes\sigma^{A_2}_{\nu}\otimes \sigma^{B_1}_{\lambda}\otimes\sigma^{B_2}_{\gamma}, \hspace{0.1cm}w_{\mu \nu \lambda \gamma} \in \mathbb {R}$, where
the set of matrices $\{\sigma^X_{\mu}\}_{\mu=0}^{d_X^2-1}$, with $\sigma^X_0=\id^X$, $\tr \sigma^X_{\mu}\sigma^X_{\nu}=d_X\delta_{\mu\nu}$, and $\tr \sigma^X_{j}=0$ for $j=1,\dots d_X^2-1$, provides a basis of ${\cal L}({\cal H}^{X})$. We refer to terms of the form $\sigma^{A_1}_{i}\otimes\id^{rest}$ ($i\geq1$) as of the type $A_1$, terms of the form $\sigma^{A_1}_{i}\otimes\sigma^{A_2}_{j}\otimes\id^{rest}$ ($i$, $j\geq1$) as of the type $A_1A_2$, and so on. In the Appendix, we prove that a matrix satisfies condition \eqref{W2} iff it contains the terms listed in this table. Each of the terms can allow signalling in at most one direction and can be realized in a situation in which either Bob's actions are not in the causal past of Alice's ($B \npreceq A$) or vice versa ($A \npreceq B$). The most general unidirectional process is a quantum channel with memory. Measurements of bipartite states that lead to non-signalling probabilities can be realized in both situations. The most general process matrix can contain terms from both rows and may not be decomposable into a mixture of quantum channels from Alice to Bob and from Bob to Alice.} \label{allowed}
\end{figure}

\begin{figure}
\begin{center}
\includegraphics[width=8.6cm]{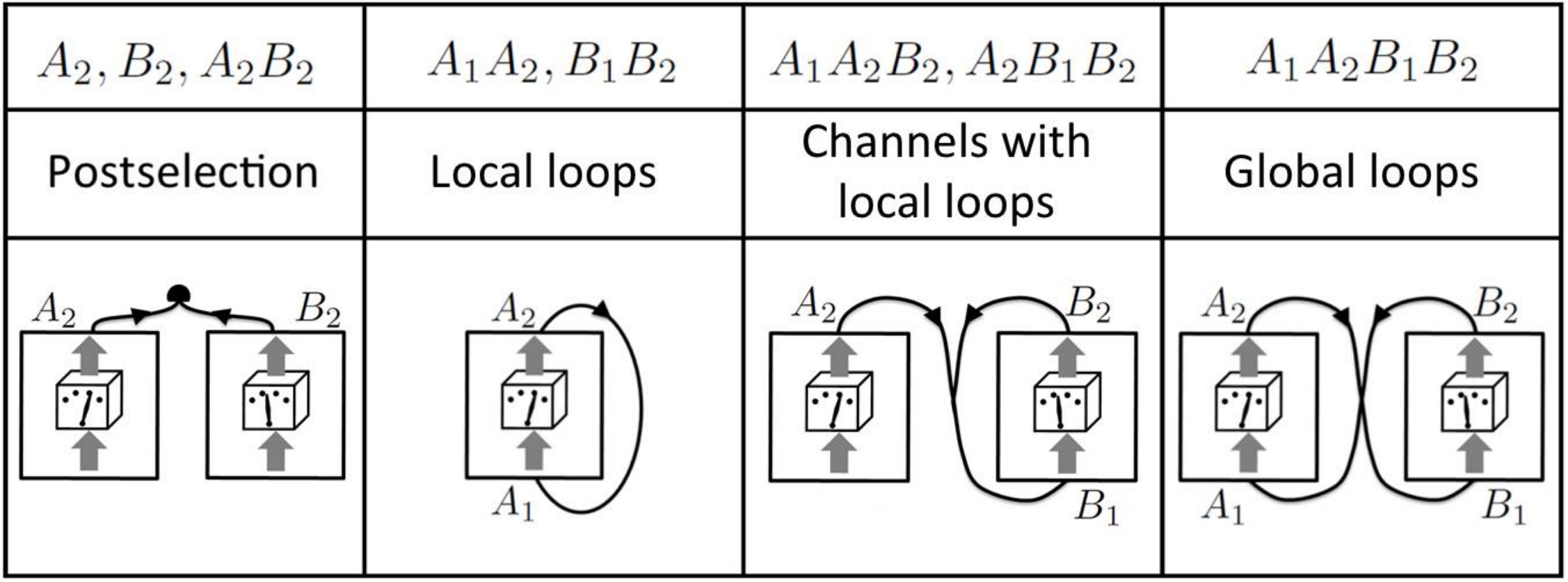}
\end{center}
\vspace{-0.5 cm}
\caption{\textbf{Terms not appearing in a process matrix.}
These terms are not compatible with local quantum mechanics because they yield non-unit probabilities for some completely positive trace-preserving maps.
A possible interpretation of these terms within our framework is that they correspond to statistical sub-ensembles of possible processes.
For example, terms of the type $A_2$ can be understood as postselection. One specific case is when a system enters a laboratory in a maximally mixed state, is subject to the map ${\cal M}$ and, after going out of the laboratory, is measured to be in some state $|\psi\ket$. The corresponding probability is given by $\tr\left[ |\psi\ket\bra\psi| {\cal M}(\frac{\id}{d})\right]$, generated in our formalism by $W^{A_1A_2}=\frac{\id}{d}^{A_1}\otimes|\psi\ket\bra\psi|^{A_2}$. Notably, correlations of the type $A_1A_2$ have been exploited in models for describing closed time-like curves \cite{Bennett2, Lloyd}. The pictures are only suggestive of the possible interpretations.} \label{notallowed}
\end{figure}

In the Appendix, we provide a complete characterization of process matrices via the terms allowed in their expansion in a Hilbert-Schmidt basis, which we relate to the possible directions of signalling they allow (see Fig.~\ref{allowed}). We also provide possible interpretations of the terms that are not allowed in a process matrix (see Fig.~\ref{notallowed} and Fig.~\ref{backwards}).

\subsection*{A causally nonseparable process}

The question whether all local quantum experiments can be embedded in a global causal structure corresponds to the question whether all process matrices are causally separable. Note that this is not a question about entanglement: all possible entangled states, and more generally all quantum circuits, correspond to matrices of the form $W^{B \npreceq A}$ or $W^{A\npreceq B}$, while the non-separable processes we are looking for cannot be written as quantum circuits or even as probabilistic mixtures of different circuits. Surprisingly, an example of such a kind exists. Consider the process matrix
\begin{equation}
\label{quantum}
	\!\!\!	W^{A_1A_2B_1B_2}=\frac{1}{4}\left[\id^{A_1A_2B_1B_2} + \frac{1}{\sqrt{2}}\left(\sigma_z^{A_2}\sigma_z^{B_1} + \sigma_z^{A_1}\sigma_x^{B_1}\sigma_z^{B_2} \right) \right],
\end{equation}
where $A_1$, $A_2$, $B_1$, and $B_2$ are two-level systems (e.g. the spin degrees of freedom of a spin-$\frac{1}{2}$ particle) and $\sigma_x$ and $\sigma_z$ are the Pauli spin matrices. It can be verified straightforwardly that conditions~\eqref{W1} and~\eqref{W2} are satisfied, hence \eqref{quantum} is a valid bipartite process. Having such a resource, Alice and Bob can play the game described above and exceed the bound on the probability of success \eqref{bound} imposed by causal order. Indeed, if Bob measures in the $z$ basis and detects one of the states $|z_{\pm}\ket$, the corresponding CJ operator contains the factor $|z_{\pm}\ket\bra z_{\pm}|^{B_1}$. Inserting this, together with Eq.~\eqref{quantum}, into the expression~\eqref{representation} for the probabilities, the term containing $\sigma_x^{B_1}$ in the process matrix is annihilated and what remains corresponds to a noisy channel from Alice to Bob. If Alice encodes her bit in the $z$ basis with the CJ operator $|z_{\pm}\ket\bra z_{\pm}|^{A_2}$, this channel allows Bob to guess Alice's bit with probability $P(y=a)=\frac{2+\sqrt{2}}{4}$. If, on the other hand, Bob measures in the $x$ basis, Eq.~\eqref{quantum} is reduced to a similar noisy channel from Bob to Alice. Bob is thus able to activate a channel in the desired direction by choosing the measurement basis (see Appendix for a detailed calculation and analysis of the protocol). In this way they can achieve
\begin{equation}
		\label{success}
		p_{succ}= \frac{2+\sqrt{2}}{4}>\frac{3}{4},
\end{equation}
which proves that~\eqref{quantum} is not causally separable. We see that, depending on his choice, Bob can effectively end up `before' or `after' Alice, each possibility with a probability $\sqrt{2}/2$.
This is remarkable, since if Alice and Bob perform their experiments inside laboratories that they believe are isolated from the outside world for the duration of their operations (e.g. by walls made of impenetrable material), and if they believe that they are able to freely choose the bits $a$, $b$, and $b'$ (e.g. by tossing a coin), they will have to conclude that the events in their experiment do not take place in a causal sequence. Indeed, the framework only assumes that the local operations from the input to the output system of each party are correctly described by quantum mechanics, and it is compatible with any physical situation in which one would have all the reasons to believe that each party's operations are freely chosen in a closed laboratory.

Interestingly, both the classical bound \eqref{bound} and the quantum violation \eqref{success} match the corresponding numbers in the CHSH-Bell inequality \cite{CHSH}, which strongly resembles inequality \eqref{bound}. However, the physical situations to which these inequalities correspond is very different: Bell inequalities can be violated in space-like separated laboratories, while \eqref{success} cannot be achieved neither with space-like nor with time-like separated laboratories. It is an open question whether \eqref{success} is the maximal possible violation allowed by quantum mechanics.

\subsection*{Classical processes are causally separable}

It is not difficult to see that if the operations of the local parties are classical, they can always be understood as taking place in a global causal structure. Classical operations can be described by transition matrices $M^{(\lambda_2 \lambda_1)}_j = P(\lambda_2,j|\lambda_1)$, where $P(\lambda_2,j|\lambda_1)$ is the conditional probability that the measurement outcome $j$ is observed and the classical output state $\lambda_2$ is prepared given that the input state is $\lambda_1$. They can be expressed in the quantum formalism as CP maps diagonal in a fixed (`pointer') basis, and the corresponding CJ operators are $M_j= \sum_{\lambda_1 \lambda_2}M^{(\lambda_2 \lambda_1)}_j |\lambda_1\ket\bra \lambda_1|^{A_1}\otimes |\lambda_2\ket\bra \lambda_2|^{A_2}$. Thus, in order to express arbitrary bipartite probabilities of classical maps, it is sufficient to consider process matrices which are diagonal in the pointer basis. In the Appendix, we provide a detailed proof that all such processes are causally separable.

\section*{\normalsize{Discussion}}

We have seen that by relaxing the assumption of definite global causal order and requiring that the standard quantum formalism holds only locally, we obtain the possibility for global causal relations that are not included in the usual formulation of quantum mechanics. The latter is reminiscent of the situation in general relativity, where by requiring that locally the geometry is that of flat Minkowski space-time, one obtains the possibility of having more general, curved space-times.

The natural question is whether ``non-causal'' quantum correlations of the kind described by our formalism can be found in nature. One can speculate that they may exist in unprobed physical regimes, such as, for example, those in which quantum mechanics and general relativity become relevant. Indeed, our result that classical theories can always be understood in terms of a global causal structure suggests the possibility that the observed causal order of space-time might not be a fundamental property of nature but rather emerge from a more fundamental theory \cite{oriti, Finkelstein, glimmers} in a quantum-to-classical transition due to, for example, decoherence \cite{Zurek} or coarse-grained measurements \cite{Kofl2007}. Once a causal structure is present, it is possible to derive relativistic space-time from it under appropriate conditions \cite{Bomb87, dari}. Furthermore, since the conformal space-time metric is a description of the causal relation between space-time points \cite{Hawking, Malament}, one can expect that an extension of general relativity to the quantum  domain would involve situations where different causal orders could coexist ``in superposition''. The formalism we presented may offer a natural route in this direction: based only on the assumption that quantum mechanics is valid locally, it yields causal relations that cannot be understood as arising from a definite, underlying order.

It is also worth noting that exotic causal structures already appear in the classical theory of general relativity. For example, there exist solutions to the Einstein equation containing closed time-like curves (CTCs) \cite{Goedel}.
In this context, it should be noted that any process matrix $W$ in our framework can be interpreted as a CPTP map from the outputs, $A_2$, $B_2$, of the parties, to their inputs, $A_1$, $B_1$. In other words, any process can be thought of as having the form of a CTC, where information is sent back in time through a noisy channel (see also Fig.~\ref{task}b). The existence of processes that do not describe definite causal order is therefore not incompatible with general relativity in principle. It is sometimes argued that CTCs should not exist since they generate logical paradoxes, such as an agent going back in time and killing his grandfather. The possible solutions that have been proposed \cite{Deutsch, Bennett2, Lloyd, Bennett1, greenberger, CTC}, in which quantum mechanics and CTCs might coexist, involve non-linear extensions of quantum theory that deviate from quantum mechanics already at the level of local experiments. Our framework, on the other hand, is by construction linear and in agreement with local quantum mechanics, and yet paradoxes are avoided, in accordance with the Novikov principle \cite{Novikov}, due to the noise in the evolution `backward in time'.

Finally we remark that instances of indefinite causal orders may also emerge in situations closer to possible laboratory implementations. As already noted, our formalism describes more general correlations than those that can be realized with a quantum circuit, that is, as a sequence of quantum gates. Recently, a new model of quantum computation which goes beyond the causal paradigm of quantum circuits by using superpositions of the `wires' connecting different gates was proposed \cite{chiribella3}. This possibility may allow breaking assumption \emph{CS} that events are localized in a causal structure. Since the instant when a system enters a device depends on how the device is wired with the rest of the computer's architecture, superpositions of wires may allow creating situations in which events are not localized in time (similarly to the way in which a quantum particle may not be localized in space). While it is an open question whether violating the causal inequality \eqref{bound} can be achieved by similar means, the present work suggests that new quantum resources for information processing might be available---beyond entanglement, quantum memories, and even `superpositions of wires'---and the formalism introduced provides a natural framework for exploring them.

\acknowledgements{We thank G. Chiribella for discussions. This work was supported by FWF projects P19570-N16 and SFB-FOQUS, FQXi, the European Commission Project Q-ESSENCE (No. 248095), and the Interuniversity Attraction Poles program of the Belgian Science Policy Office, under grant IAP P6-10 $\ll$photonics@be$\gg$. O.O. acknowledges the support of the European Commission under the Marie Curie Intra-European Fellowship Programme (PIEF-GA-2010-273119).  F.C. is a member of the FWF Doctoral Program CoQuS (W 1210).}

\section*{\normalsize{Appendix}}

\subsection{{Formal derivation of the causal inequality}}

A causal structure (for instance, space-time) is a set of event locations equipped with a partial ordering relation $\preceq$ that defines the possible causal relations between events at these locations. If $A$ and $B$ are two such locations, $A\preceq B$ reads ``$A$ is in the \textit{causal past} of $B$'', or equivalently, ``$B$ is in the \textit{causal future} of $A$'' (e.g. if $A$ and $B$ are space-time points, $A\preceq B$ corresponds to $A$ being in the past light cone of $B$). Operationally, if $A\preceq B$, an agent at $A$ can \textit{signal} to an agent at $B$ by encoding information in events at $A$ that get correlated with events at $B$ which the other agent can observe. (Formally, signalling from $A$ to $B$ is the existence of statistical correlations between a random variable at $A$ which can be chosen freely, and another random variable at $B$. By definition, a freely chosen variable is one that can be correlated only with variables in its causal future. Note that a freely chosen variable is an idealization since the result of a coin toss or any other candidate for a freely chosen variable may be correlated with initial conditions in the past or with space-like separated events, but these correlations are ignored as not relevant to the variables of interest.) The fact that the relation $\preceq$ is a partial order means that it satisfies the following conditions: 1) $A\preceq A$ (reflexivity); 2) if $A\preceq B$ and $B\preceq C$, then $A\preceq C$ (transitivity); and 3) if $A\preceq B$ and $B\preceq A$, then $A=B$ (antisymmetry). The last condition says that if $A$ and $B$ are two different locations, there can either be signalling from $A$ to $B$, or vice versa, but no signalling in both directions is possible (i.e. there are no causal loops). If $A$ is \textit{not} in the causal past of $B$, we will write $A\npreceq B$. Note that in a causal structure both $A\npreceq B$ and $B\npreceq A$ may hold (as in the case when $A$ and $B$ are space-like separated), and at least one of the two \textit{must} hold for $A\neq B$. We will denote the situation where both $A\npreceq B$ and $B\npreceq A$ hold by $A\npreceq\nsucceq B$.

Since every event specifies an event location, we will use the same notation directly for events. For instance, if $X$ and $Y$ are two events such that the location of $X$ is in the causal past of the location of $Y$, we will write $X\preceq Y$ (similarly for $\npreceq$ and $\npreceq\nsucceq$).

The main events in our communication task are the systems entering Alice's and Bob's laboratories, which we will denote by $A_1$ and $B_1$, respectively, and the parties producing the bits $a$, $b$, $b'$, $x$, and $y$, which we will denote by the same letters as the corresponding bits. The fact that Alice generates the bit $a$ and produces her guess $x$ \textit{after} the system enters her laboratory means that $A_1\preceq a,y$. Similarly, we have $B_1 \preceq b', b, y$.

The assumptions behind the causal inequality are:

\emph{Causal structure} (\emph{CS})---The events $A_1$, $B_1$, $a$, $b$, $b'$, $x$, $y$ are localized in a causal structure.

\emph{Free choice} (\emph{FC})---Each of the bits $a$, $b$, and $b'$ can only be correlated with events in its causal future (this concerns only events relevant to the task). We assume also that each of them takes values $0$ or $1$ with probability $1/2$.

\emph{Closed laboratories} (\emph{CL})---$x$ can be correlated with $b$ only if $b\preceq A_1$, and $y$ can be correlated with $a$ only if $a\preceq B_1$.

We want to show that these assumptions imply
\begin{gather}
p_{succ}=\frac{1}{2}p(x=b|b'=0)+\frac{1}{2}p(y=a|b'=1)\leq \frac{3}{4}
\end{gather}
for the success probability that Alice and Bob can achieve in their task.

First, notice that assumption \emph{FC} implies that the bits $a$, $b$, and $b'$ are independent of each other (\emph{CS} is assumed throughout). Indeed, there are two general ways in which the three bits could be correlated---two of them are correlated with each other while the third one is independent, or each of them is correlated with the other two. In the first case, the free-choice assumption implies that the two correlated bits would have to be in each other's causal pasts, which is impossible. In the second case, each of the bits would have to be in the causal past of the other two, which is again impossible. Hence, the bits are uncorrelated.

Next, consider the following three possibilities that can be realized in a causal structure (\emph{CS} is assumed throughout): $A_1\preceq B_1$, $B_1\preceq A_1$, $A_1\npreceq\nsucceq B_1$. Since these possibilities are mutually exclusive and exhaustive, their probabilities satisfy $p (A_1\preceq B_1)+ p(B_1\preceq A_1)+p(A_1\npreceq\nsucceq B_1)= 1$. From assumption \emph{FC} it follows that the bits $a$, $b$, and $b'$ are independent of which of these possibilities is realized. To see this, consider for instance $b'$. Since $B_1\preceq b'$, we have that $b'$ must be independent of whether $A_1$ takes place in the causal past of $B_1$ or not, i.e. $p(A_1\preceq B_1 |b') = p(A_1\preceq B_1)$. Similarly, $b'$ must be independent of whether $A_1$ takes place in the larger region which is a complement of the causal future of $B_1$, which implies $p(B_1\npreceq A_1 |b') = p(B_1\npreceq A_1)$. But $p(B_1\npreceq A_1 |b')= p(A_1\preceq B_1 |b')+p(A_1\npreceq\nsucceq B_1|b') = p(A_1\preceq B_1)+p(A_1\npreceq\nsucceq B_1|b')$, while $p(B_1\npreceq A_1)= p(A_1\preceq B_1)+p(A_1\npreceq\nsucceq B_1)$, which implies $p(A_1\npreceq\nsucceq B_1|b')= p(A_1\npreceq\nsucceq B_1)$. Finally, since $p(A_1\preceq B_1 |b')+ p(A_1\npreceq\nsucceq B_1|b')+ p(B_1\preceq A_1 |b') = p(A_1\preceq B_1)+ p(A_1\npreceq\nsucceq B_1)+ p(B_1\preceq A_1 |b') =1 = p(A_1\preceq B_1)+ p(A_1\npreceq\nsucceq B_1)+ p(B_1\preceq A_1)$, we have $p(B_1\preceq A_1|b')=p(B_1\preceq A_1)$. An analogous argument shows that $a$ and $b$ are also independent of the causal relation between $A_1$ and $B_1$.

Using the above, the success probability can be written
\begin{widetext}
\begin{gather}
p_{succ}=\frac{1}{2}p(x=b|b'=0)+\frac{1}{2}p(y=a|b'=1)\notag\\
=\frac{1}{2}p(x=b|b'=0;A_1\preceq B_1)p(A_1\preceq B_1)+\frac{1}{2}p(x=b|b'=0;B_1\preceq A_1)p(B_1\preceq A_1) +  \frac{1}{2}p(x=b|b'=0;A_1\npreceq\nsucceq B_1)p(A_1\npreceq\nsucceq B_1)\notag\\
+
\frac{1}{2}p(y=a|b'=1;A_1\preceq B_1)p(A_1\preceq B_1)+\frac{1}{2}p(y=a|b'=1;B_1\preceq A_1)p(B_1\preceq A_1)+\frac{1}{2}p(y=a|b'=1;A_1\npreceq\nsucceq B_1)p(A_1\npreceq\nsucceq B_1)\notag\\
= \left(\frac{1}{2}p(x=b|b'=0;A_1\preceq B_1)+\frac{1}{2}p(y=a|b'=1;A_1\preceq B_1)\right)p(A_1\preceq B_1)\notag\\
+\left(\frac{1}{2}p(x=b|b'=0;B_1\preceq A_1) +\frac{1}{2}p(y=a|b'=1;B_1\preceq A_1)\right)p(B_1\preceq A_1)\notag\\
+\left(\frac{1}{2}p(x=b|b'=0;A_1\npreceq\nsucceq B_1)+\frac{1}{2}p(y=a|b'=1;A_1\npreceq\nsucceq B_1)\right)p(A_1\npreceq\nsucceq B_1).\label{prelim}
\end{gather}
\end{widetext}
If $A_1\preceq B_1$ (which implies $B_1\npreceq A_1$), from the transitivity of partial order it follows that $A_1\preceq b$ (and thus $b\npreceq A_1$). From assumption \emph{CL}, $x$ can only be correlated with $b$ if $b$ is in the causal past of $A_1$, thus $p(b| x; A_1\preceq B_1)= p(b| A_1\preceq B_1)=\frac{1}{2}$ [the last equality follows from the independence of $b$ from the causal relations between $A_1$ and $B_1$, together with assumption \emph{FC}]. Using also that $b$ and $b'$ are independent, we thus obtain $p(x=b|b'=0;A_1\preceq B_1)= p(b=0; x=0|b'=0;A_1\preceq B_1)+p(b=1,x=1|b'=0;A_1\preceq B_1) = p(b=0| x=0; b'=0;A_1\preceq B_1)p(x=0| b'=0;A_1\preceq B_1)+p(b=1|x=1; b'=0;A_1\preceq B_1) p( x=1| b'=0;A_1\preceq B_1) = \frac{1}{2} p(x=0| b'=0;A_1\preceq B_1)+ \frac{1}{2} p ( x=1| b'=0;A_1\preceq B_1) = \frac{1}{2}$.

If $B_1\preceq A_1$ (which implies $A_1\npreceq B_1$), by an analogous argument we obtain $p(y=a|b'=1;B_1\preceq A_1)=\frac{1}{2}$. Finally, if $A_1\npreceq\nsucceq B_1$, we have both $p(y=a|b'=1;A_1\npreceq\nsucceq B_1)=\frac{1}{2}$ and $p(x=b|b'=0;A_1\npreceq\nsucceq B_1)=\frac{1}{2}$.
Substituting this in Eq.~\eqref{prelim}, we obtain

\begin{gather}
p_{succ}
= \left(\frac{1}{4}+\frac{1}{2}p(y=a|b'=1;A_1\preceq B_1)\right)p(A_1\preceq B_1)\notag\\
+\left(\frac{1}{2}p(x=b|b'=0;B_1\preceq A_1) +\frac{1}{4}\right)p(B_1\preceq A_1)\notag\\
+\left(\frac{1}{4}+\frac{1}{4}\right)p(A_1\npreceq\nsucceq B_1)\notag\\
\leq \frac{3}{4}p(A_1\preceq B_1)+ \frac{3}{4} p(B_1\preceq A_1)+  \frac{3}{4}p(A_1\npreceq\nsucceq B_1)=\frac{3}{4}.\
\end{gather}
This completes the proof.

\subsection{{Definition of process matrices}}

In this section we will derive the linear representation \eqref{representation} as well as the conditions \eqref{W1} and \eqref{W2} that a process matrix has to satisfy.

\emph{Linearity of probabilities.} A quantum instrument  \cite{instrument} is defined as a set $\{{\cal M}_j\}_{j=1}^n$ of CP maps such that ${\cal M}=\sum_{j=1}^n{\cal M}_j$ is a CPTP map.
Our main assumption is that the description of the operations in the individual laboratories is in agreement with quantum mechanics.
In particular, we derive linearity from the quantum mechanical representation of probabilistic mixtures and of coarse-graining of operations.
Consider first an instrument $\{\tilde{{\cal M}}_j\}_{j=1}^n$ defined as the randomization of two different instruments $\{{\cal M}_j\}_{j=1}^n$ and $\{{\cal N}_j\}_{j=1}^n$, where the first is performed with probability $p$ and the second with probability $(1-p)$. The probability to observe the outcome $j$ is, by definition, $P(\tilde{{\cal M}}_j)=pP({\cal M}_j)+(1-p)P({\cal N}_j)$. In quantum mechanics randomization is described as a convex linear combination, $\tilde{{\cal M}}_j=p{\cal M}_j + (1-p){\cal N}_j$. We can then conclude that the probability must respect linear convex combinations: $P\left(p{\cal M}_j + (1-p){\cal N}_j \right) = pP({\cal M}_j)+(1-p)P({\cal M}_j)$.
Consider then the coarse-graining of an instrument $\{{\cal M}_j\}_{j=1}^{n}$. This is realized when two or more outcomes, for example those corresponding to the labels $j=n-1$ and $j=n$, are treated as a single one. In the resulting instrument $\{\tilde{{\cal M}}_j\}_{j=1}^{n-1}$ all non coarse-grained outcomes correspond to the original CP maps $\tilde{{\cal M}}_j = {\cal M}_j$ for $j=1,\dots n-2$, while the probability of the coarse-grained outcome is given by $P(\tilde{{\cal M}}_{n-1})=P({\cal M}_{n-1}) + P({\cal M}_n)$. In quantum mechanics, the CP map corresponding to the coarse graining of two outcomes is represented by the sum of the respective CP maps, $\tilde{{\cal M}}_{n-1} = {\cal M}_{n-1}+{\cal M}_n$, from which it follows that $P\left({\cal M}_{n-1}+{\cal M}_n\right)=P\left({\cal M}_{n-1}\right)+P\left({\cal M}_n\right)$. Randomization and coarse graining together impose linearity.
The argument can be repeated for two (or more) parties, yielding the conclusion that all bipartite probabilities compatible with a local quantum mechanical description are bilinear functions, $P\left({\cal M}^A_i,{\cal M}^B_j\right)=\omega\left({\cal M}^A_i,{\cal M}^B_j\right)\in[0,1]$, of the local CP and trace-nonincreasing maps ${\cal M}^A_i$, ${\cal M}^B_j$.

Thanks to the CJ isomorphism, it is possible to represent bilinear functions of CP maps as bilinear functions of matrices: $\omega \leftrightarrow \tilde{\omega}:{\cal L}({\cal H}^{A_1}\otimes{\cal H}^{A_2})\times{\cal L}({\cal H}^{B_1}\otimes{\cal H}^{B_2})\rightarrow \mathbb {R}$. In general, multilinear functions on a set of vector spaces $V^1\times V^2\times\dots$ are isomorphic to linear functions on $V^1\otimes V^2\otimes\dots$, hence the probabilities can be written as linear functions on ${\cal L}({\cal H}^{A_1}\otimes{\cal H}^{A_2}\otimes{\cal H}^{B_1}\otimes{\cal H}^{B_2})$. Using the Hilbert-Schmidt scalar product, we can identify each real linear function with an element of the same space, $\tilde{\omega} \leftrightarrow W^{A_1A_2B_1B_2}\in {\cal L}({\cal H}^{A_1}\otimes{\cal H}^{A_2}\otimes{\cal H}^{B_1}\otimes{\cal H}^{B_2})$, arriving at the representation \eqref{representation}.

\emph{Nonnegativity and normalization of probabilities.} The requirement that the probabilities are non-negative for any pair of CP maps $\mathcal{M}^A$ and $\mathcal{M}^B$ imposes the restriction that $W$ is positive on pure tensors (POPT) \cite{popt} with respect to the partition $A_1A_2-B_1B_2$. These are matrices such that
\begin{equation}
		\label{popt}
		\begin{split}
		\tr \left[W^{A_1A_2B_1B_2}\left(M^{A_1A_2}\otimes M^{B_1B_2}\right)\right] \geq 0, \\
		\forall M^{A_1A_2}\geq 0 ,M^{B_1B_2} \geq 0.
		\end{split}
\end{equation}
The condition has to be imposed for arbitrary positive semidefinite matrices $M^{A_1A_2}$ and $M^{B_1B_2}$ because these are the CJ matrices of CP maps.

We additionally assume that the parties can share arbitrary (possibly entangled) ancillary states independent of the process, and use them in their local operations. The latter means that each party can extend the input space of her/his operations to the ancillas, which we denote by $A'_1$ and $B'_1$ for Alice and Bob, respectively, and apply arbitrary quantum operations with CP maps $\mathcal{M}^A:\mathcal{L}(\mathcal{H}^{A_1'}\otimes \mathcal{H}^{A_1})\rightarrow \mathcal{L}(\mathcal{H}^{A_2})$, $\mathcal{M}^B:\mathcal{L}(\mathcal{H}^{B_1'}\otimes \mathcal{H}^{B_1})\rightarrow \mathcal{L}(\mathcal{H}^{B_2})$. (One can similarly extend the output systems, but this is not necessary for our argument.) The assumption that the ancillary systems contain a joint quantum state independent of the process means that if separate operations are applied on the ancillas and the original systems, the joint probability distribution for the outcomes is a product of two distributions---one for the outcomes on the ancillas, which is the same as one arising from a measurement on a quantum state $\rho^{A'_1B'_1}$, and another one for the outcomes on the original systems, which is given by Eq.~\eqref{representation} with the original $W^{A_1A_2B_1B_2}$. These requirements imply that the extended process matrix is given by $W^{A_1'A_1A_2B_1'B_1B_2}=\rho^{A'_1B'_1}\otimes W^{A_1A_2B_1B_2}$. If we then require that the probabilities for extended operations are non-negative, one has
\begin{gather}\label{popt2}
		\tr \left[\rho^{A'_1B'_1}\otimes W^{A_1A_2B_1B_2}\left(M^{A'_1A_1A_2}\otimes M^{B'_1B_1B_2}\right)\right] \geq 0, \\\nonumber
		 \forall M^{A'_1A_1A_2}, M^{B'_1B_1B_2}, \rho^{A'_1B'_1} \geq 0.
\end{gather}
It was shown \cite{popt} that condition \eqref{popt2} is satisfied if and only if $W^{A_1A_2B_1B_2}$ is positive semidefinite (a class strictly smaller than POPT), which is condition \eqref{W1}.

Additionally, probabilities must be normalized: $1=\sum_{ij}\omega\left({\cal M}^A_i,{\cal M}^B_j\right)=\omega\left(\sum_i{\cal M}^A_i,\sum_j{\cal M}^B_j\right)$, which means
\begin{equation}
		\label{omega}
		\omega\left({\cal M}^A,{\cal M}^B\right) = 1, \;\; \forall\, \mbox{CPTP } {\cal M}^A,\, {\cal M}^B.
\end{equation}
Condition \eqref{W2} can be deduced from Eq.~\eqref{omega} simply by noticing that for a CPTP map ${\cal M}$ the corresponding CJ matrix satisfies the condition $\tr_{A_2} M^{A_1A_2} = \tr_{A_2} \left({\cal I}\otimes {\cal M}(|\phi^+\ket\bra \phi^+|) \right)^{\mathrm T}= \left[\tr_{A_2} \left({\cal I}\otimes {\cal M}(|\phi^+\ket\bra \phi^+|)\right)\right]^{\mathrm T} = \id^{A_1}$. To see that this is also a sufficient condition for a map to be trace-preserving, it is enough to consider the inverse direction of the CJ isomorphism,
\begin{equation}
		\label{inverse}
		{\cal M}(\rho^{A_1}) := \left(\tr_{A_1}\left[\rho^{A_1} M^{A_1A_2}\right]\right)^{\mathrm T}.
\end{equation}

\subsection{{Characterization of process matrices}}

Here we derive necessary and sufficient conditions for a matrix $W^{A_1A_2B_1B_2}$ to satisfy Eq.~\eqref{W1} and Eq.~\eqref{W2} in terms of an expansion of the matrix in a Hilbert-Schmidt basis. A Hilbert-Schmidt basis of ${\cal L}({\cal H}^{X})$ is given by a set of matrices $\{\sigma^X_{\mu}\}_{\mu=0}^{d_X^2-1}$, with $\sigma^X_0=\id_X$, $\tr \sigma^X_{\mu}\sigma^X_{\nu}=d_X\delta_{\mu\nu}$, and $\tr \sigma^X_{j}=0$ for $j=1,\dots d_X^2-1$. A general element of ${\cal L}({\cal H}^{A_1}\otimes {\cal H}^{A_2}\otimes{\cal H}^{B_1}\otimes {\cal H}^{B_2})$ can be expressed as
\begin{gather}
W^{A_1A_2B_1B_2}= \sum_{\mu \nu \lambda \gamma} w_{\mu \nu \lambda \gamma} \sigma^{A_1}_{\mu}\sigma^{A_2}_{\nu}\sigma^{B_1}_{\lambda}\sigma^{B_2}_{\gamma}, \hspace{0.3cm}w_{\mu \nu \lambda \gamma} \in \mathbb {C}
\end{gather}
(we omit tensor products and identity matrices whenever there is no risk of confusion). Since a process matrix has to be Hermitian, we consider only the cases
\begin{gather}
\hspace{0.3cm}w_{\mu \nu \lambda \gamma} \in \mathbb {R}.
\end{gather}

We will refer to terms of the form $\sigma^{A_1}_{i}\otimes\id^{rest}$ ($i\geq1$) as of the type $A_1$, terms such as $\sigma^{A_1}_{i}\otimes\sigma^{A_2}_{j}\otimes\id^{rest}$ ($i$, $j\geq1$) as of the type $A_1A_2$, and so on. The properties of a process matrix can be analysed with respect to the terms it contains. For example, terms of the type $A_1B_1$ produce non-signalling correlations between the measurements, terms such as $A_2B_1$ correlate Alice's outputs with Bob's inputs, yielding signalling from Alice to Bob, etc., as illustrated in Fig.~\ref{allowed}. Note that not all terms are compatible with the condition~\eqref{W2}. We will prove that a matrix $W$ satisfies condition~\eqref{W2} if and only if it only contains the terms listed in Fig.~\ref{allowed}.

The CJ matrix of a local operation can be similarly written $M^{X_1X_2}= \sum_{\mu \nu} r_{\mu \nu} \sigma^{X_1}_{\mu}\sigma^{X_2}_{\nu}$, $r_{\mu \nu} \in \mathbb {R}$. The condition $\tr_{X_2} M^{X_1X_2} = \id^{X_1}$ is equivalent to the requirement $r_{00} = \frac{1}{d_{X_2}}$, $r_{i0}=0$ for $i>0$. Thus CJ matrices corresponding to CPTP maps have the form
\begin{gather}
\label{CPTP}
M^{X_1X_2} = \frac{1}{d_{X_2}}\left(\id+\sum_{i>0}a_{i}\sigma^{X_2}_{i}+ \sum_{ij>0}t_{ij}\sigma^{X_1}_{i}\sigma^{X_2}_{j}\right), \\\notag
a_{i}, t_{ij} \in \mathbb {R}.
\end{gather}

Let us consider first the case of a single party, say, Alice. Since the set of matrices $M^{A_1A_2}\geq 0$ is a substantial set, condition \eqref{W2} can be equivalently imposed on arbitrary matrices of the form \eqref{CPTP} and, for a single party, it can be rewritten as
\begin{gather} \notag
	\frac{1}{d_{A_2}} \tr \left[W^{A_1A_2}\left(\id+\sum_{i>0}a_{i}\sigma^{A_2}_{i} + \sum_{ij>0}t_{ij}\sigma^{A_1}_{i}\sigma^{A_2}_{j}\right)\right] = 1, \\\notag
	\forall \;\; a_{i}, t_{ij} \in \mathbb {R}.
\end{gather}
Using an expansion of the process matrix in the same basis in a similar way, $W^{A_1A_2}= \sum_{\mu \nu} w_{\mu \nu} \sigma^{A_1}_{\mu}\sigma^{A_2}_{\nu}$, $w_{\mu \nu} \in \mathbb {R}$, the above condition becomes
\begin{gather} \notag
	d_{A_1}\left(w_{00} + \sum_{i>0}w_{0i}a_{i} + \sum_{ij>0}w_{ij}t_{ij} \right) = 1, \\\notag
	\forall \;\; a_{i}, t_{ij} \in \mathbb {R},
\end{gather}
and one obtains $w_{00}=\frac{1}{d_{A_1}}$, $w_{0i}=w_{ij}=0$ for $i, j >0$. Thus the most general process matrix observed by a single party has the form
\begin{gather}\label{singleW}
W^{A_1A_2} = \frac{1}{d_{A_1}}\left(\id+\sum_{i>0}v_{i}\sigma^{A_1}_{i}\right), \\\notag
v_{i} \in \mathbb {R}, \;\; W^{A_1A_2}\geq 0,
\end{gather}
which can be recognized as a state. This result---that all probabilities a single agent can observe are described by quantum states---is an extension of Gleason's theorem from POVMs  \cite{Gleason, CFMR} to CP maps (note that here the linear structure of quantum operations is assumed, while in Gleason's theorem for POVMs it is derived from different hypotheses. However, by a similar argument one could derive linearity for CP maps too).

Let us now consider a bipartite process matrix, $W^{A_1A_2B_1B_2}= \sum_{\mu \nu \lambda \gamma} w_{\mu \nu \lambda \gamma} \sigma^{A_1}_{\mu}\sigma^{A_2}_{\nu}\sigma^{B_1}_{\lambda}\sigma^{B_2}_{\gamma}$, $w_{\mu \nu \lambda \gamma} \in \mathbb {R}$. We have to impose \eqref{W2} for arbitrary matrices $M^{A_1A_2}$, $M^{B_1B_2}$ of the form \eqref{CPTP}. First, if we fix $M^{B_1B_2}=\frac{\id^{B_1B_2}}{d_{B_2}}$, we obtain
\begin{gather} \notag
	d_{A_1}d_{B_1}\left(w_{0000} + \sum_{i>0}w_{0i00}a_{i} + \sum_{ij>0}w_{ij00}t_{ij} \right) = 1 \\\notag
	\forall \;\; a_{i}, t_{ij} \in \mathbb {R},
\end{gather}
which imposes $w_{0000}=\frac{1}{d_{A_1}d_{B_1}}$ and $w_{0i00}=w_{ij00}=0$ for $i, j >0$. Similarly, by fixing $M^{A_1A_2}=\frac{\id^{A_1A_2}}{d_{A_2}}$, we can derive $w_{000i}=w_{00ij}=0$ for $i, j >0$. Finally, imposing \eqref{W2} for arbitrary
\begin{gather} \notag
	M^{A_1A_2} = \frac{1}{d_{A_2}}\left(\id+\sum_{i>0}a_{i}\sigma^{A_2}_{i}+ \sum_{ij>0}t_{ij}\sigma^{A_1}_{i}\sigma^{A_2}_{j}\right), \\\notag
	M^{B_1B_2} = \frac{1}{d_{B_2}}\left(\id+\sum_{k>0}b_{k}\sigma^{B_2}_{k}+ \sum_{kl>0}s_{kl}\sigma^{B_1}_{k}\sigma^{B_2}_{l}\right),
\end{gather}
we obtain
\begin{align*}
	  & \sum_{ik>0} w_{0i0k}a_ib_k + \sum_{ikl>0} w_{0ikl}a_is_{kl} \\
	+	& \sum_{ijk>0} w_{ij0k}t_{ij}b_k + \sum_{ijkl>0} w_{ijkl}t_{ij}s_{kl} = 0, \\
	   \forall \; & a_{i}, t_{ij}, b_k, s_{kl} \in \mathbb {R},
\end{align*}
from which we conclude that the most general matrix that satisfies \eqref{W2} has the form
\begin{align*}
	W^{A_1A_2B_1B_2} 	&= \frac{1}{d_{A_1}d_{B_1}}\left(\id + \sigma^{B\preceq A}+ \sigma^{A\preceq B}+\sigma^{A\npreceq\nsucceq B}\right) ,\\
	\sigma^{B\preceq A}	&:= \sum_{ij>0}c_{ij}\sigma^{A_1}_{i}\sigma^{B_2}_{j} + \sum_{ijk>0}d_{ijk}\sigma^{A_1}_{i}\sigma^{B_1}_{j}\sigma^{B_2}_{k},\\
	\sigma^{A\preceq B}	&:= \sum_{ij>0}e_{ij}\sigma^{A_2}_{i}\sigma^{B_1}_{j} + \sum_{ijk>0}f_{ijk}\sigma^{A_1}_{i}\sigma^{A_2}_{j}\sigma^{B_1}_{k},\\
	\sigma^{A\npreceq\nsucceq B} 	&:= \sum_{i>0}v_{i}\sigma^{A_1}_{i} + \sum_{i>0}x_{i}\sigma^{B_1}_{i} + \sum_{ij>0}g_{ij}\sigma^{A_1}_{i}\sigma^{B_1}_{j},\\
	\mbox{where }&	c_{ij}, d_{ijk}, e_{ij}, f_{ijk}, g_{ij}, v_{i}, x_{i} \in \mathbb {R}.
\end{align*}
This form, together with the condition $W^{A_1A_2B_1B_2}\geq 0$, completely characterizes the most general bipartite process matrix.


\subsection{{Terms not appearing in process matrices}}

\begin{figure}
\begin{center}
\includegraphics[width=3.4cm]{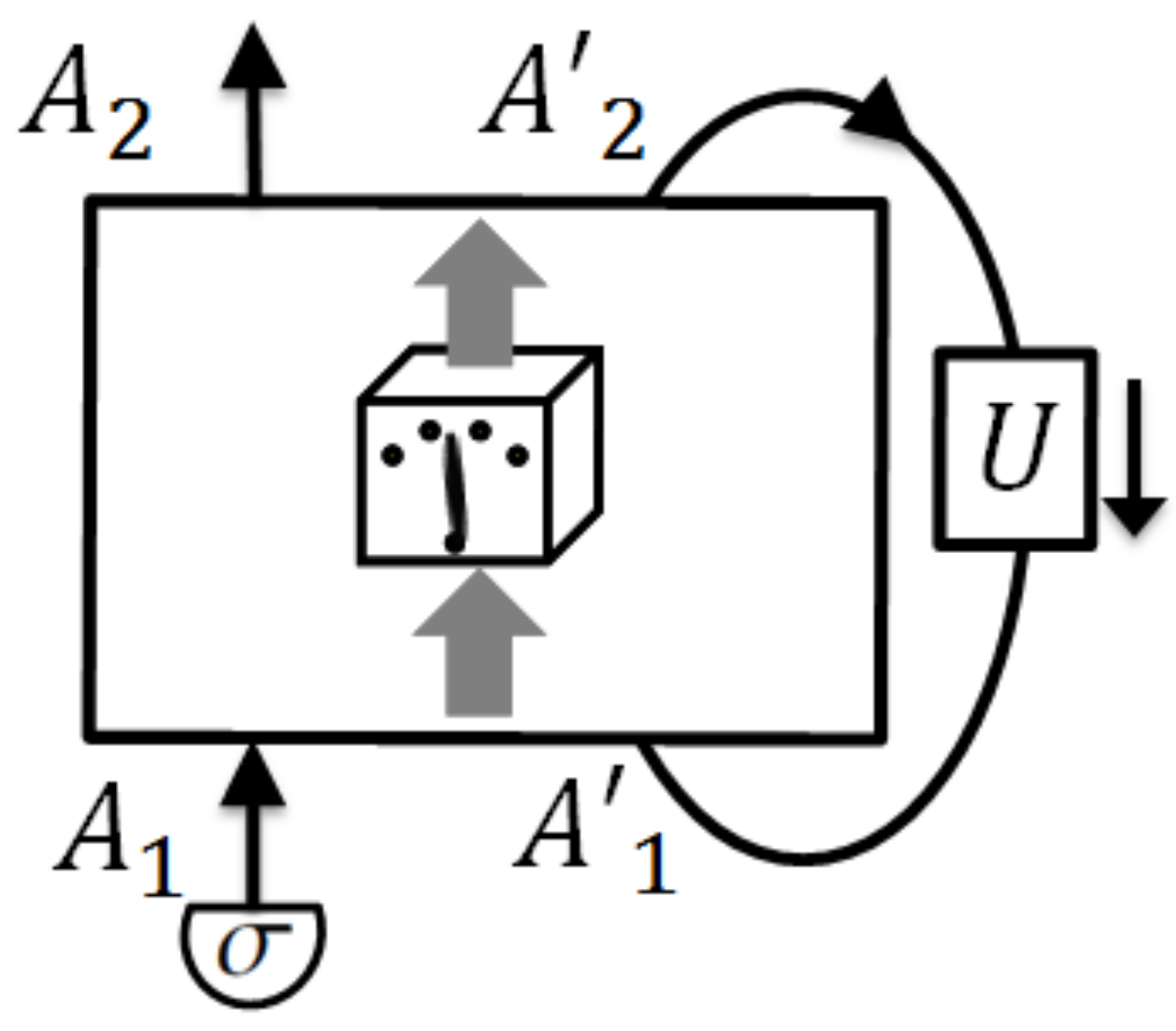}
\end{center}
\caption{\textbf{Nonlinear model of closed time-like curve.} In the model of closed time-like curves considered in Refs.~\cite{Bennett2, Lloyd}, a chronology-respecting system $A$, initially in a state $\sigma$, interacts with a second system, $A'$, which travels back in time according to a unitary $U$. This model can be represented in our formalism by an `unphysical' process matrix, i.e. one for which probabilities do not sum up to one.} \label{backwards}
\end{figure}

The not-allowed terms are listed in Fig.~\ref{notallowed}, along with possible interpretations. Particularly interesting are the cases involving terms of the type $A_1A_2$. These would correlate Alice's output with her input and not give unit probabilities for some CPTP maps that she can choose to perform. This kind of correlations resemble a `backward in time' transmission of information: one can imagine that they can be generated by a quantum channel `in the inverse order', from the output $A_2$ to the input $A_1$. It is worth noting that a recently proposed model of closed time-like curves \cite{Bennett2, Lloyd} can be expressed precisely in this way. Using our terminology, such a model considers an agent receiving two quantum systems in her laboratory: a chronology-respecting system $A$ and a second system $A'$ which, after leaving the laboratory, is sent back in time to the laboratory's entrance (see Fig.~\ref{backwards}). This can be described by the process matrix $W^{A_1A_1'A_2A_2'} = \sigma^{A_1}\otimes\id^{A_2}\otimes \left( U\otimes \id |\phi^+\ket\bra\phi^+|^{A_1'A_2'} U^{\dag} \otimes \id  \right)$, where $\sigma^{A_1}$ is the state of the chronology-respecting system when it enters the laboratory and $\left( U\otimes \id |\phi^+\ket\bra\phi^+|^{A_1'A_2'} U^{\dag} \otimes \id  \right)$ is a process matrix corresponding to a unitary $U$ from $A_2'$ to $A_1'$, describing the evolution back in time of the chronology-violating system.
(The labels $A_1$, $A_1'$ represent the two systems entering the laboratory, while $A_2$, $A_2'$ represent the systems going out. Note that here the two systems belong to the same laboratory and they can undergo any joint operation.) In this model, probabilities have to be renormalized in order to sum up to one, which introduces a non-linearity that violates our original assumptions (in particular, as opposed to quantum mechanics, probabilities are contextual in this model, since it is necessary to specify the events that did not occur in order to perform the renormalization step). The same can be said for Deutsch's model of closed time-like curves \cite{Deutsch}, which is also non-linear (although it uses a different mechanism to obtain well-defined probabilities) and thus violates our premise that ordinary quantum mechanics holds locally in each laboratory.

\subsection{{Violation of the causal inequality}}

The process described by Eq.~\eqref{quantum} can be exploited for the task described above in the following way. Alice always measures the incoming qubit in the $z$ basis, assigning the value $x=0$ to the outcome $|z_+\ket$ and $x=1$ to $|z_-\ket$. She then reprepares the qubit, encoding $a$ in the same basis, and sends it away. It is easy to see that the CP map corresponding to the detection of a state $|\psi\ket$ and repreparation of another state $|\phi\ket$ has CJ matrix $|\psi\ket\bra \psi|^{A_1}\otimes|\phi\ket\bra \phi|^{A_2}$. Accordingly, the possible operations performed by Alice can be represented compactly by the CJ matrix
\begin{equation}
		\label{alice}
		\xi^{A_1A_2}(x,a)=\frac{1}{4} \left[\id+(-1)^x\sigma_z \right]^{A_1}\otimes\left[\id+(-1)^a\sigma_z \right]^{A_2}.
\end{equation}
Bob adopts the following protocol. If he wants to read Alice's bit ($b'=1$), he measures the incoming qubit in the $z$ basis and assigns $y=0$, $y=1$ to the outcomes $|z_+\ket$, $|z_-\ket$, respectively (the repreparation is unimportant in this case). If he wants to send his bit ($b'=0$), he measures in the $x$ basis and, if the outcome is $|x_+\ket$, he encodes $b$ in the $z$ basis of the outgoing qubit as $0\rightarrow |z_+\ket$, $1\rightarrow |z_-\ket$, while, if the outcome is $|x_-\ket$, he encodes it as $0\rightarrow |z_-\ket$, $1\rightarrow |z_+\ket$. The CJ matrix representing Bob's CP map is
\begin{align}
	\label{bob}
		                   \eta^{B_1B_2}(y,b,b')&=  b'\eta_1^{B_1B_2}(y,b)+(b'\oplus 1)\eta_2^{B_1B_2}(y,b),  \\
	\label{bobreceive}	 \eta_1^{B_1B_2}(y,b) &=  \frac{1}{2} \left[\id + (-1)^y\sigma_z  \right]^{B_1}\otimes\rho^{B_2},  \\
	\label{bobsend}			 \eta_2^{B_1B_2}(y,b) &= \frac{1}{4}  \left[\id + (-1)^y\sigma_x  \right]^{B_1}\otimes\left[\id + (-1)^{b+y}\sigma_z\right]^{B_2},
\end{align}
where $\rho^{B_2}$ is the arbitrary state prepared when $b'=1$ (with $\tr \rho^{B_2} =1$) and $\oplus$ denotes the sum modulo 2. Note that in Eq. \eqref{bobsend} Bob's assignment $|x_+\ket\rightarrow y=0$, $|x_-\ket\rightarrow y=1$ for the outcome of his measurement is arbitrary since for $b'=0$ he is not trying to correlate $y$ with $a$.

The probabilities for different possible outcomes, when the described protocol is applied to the process \eqref{quantum}, are given, according to \eqref{representation}, by $P(xy|abb')= \tr \left[W^{A_1A_2B_1B_2}\left(\xi^{A_1A_2}(x,a)\eta^{B_1B_2}(y,b,b')\right)\right]$. In order to calculate the success probability, we need as intermediate steps $P(y|ab,b'=1)=\sum_{x} P(xy|ab,b'=1)$ and $P(x|ab,b'=0)=\sum_{y} P(xy|ab,b'=0)$. Notice that when the outcome of one party is ignored, it is always possible to identify a specific state in which the other party receives the qubit. For example, to average out Alice's outcomes one has to calculate
\begin{displaymath}
\begin{split}
	&\sum_x \tr \left[W^{A_1A_2B_1B_2}\left(\xi^{A_1A_2}(x,a)\eta^{B_1B_2}(y,b,b')\right)\right] \\
	&= \tr_{B_1B_2}\left\{ \eta^{B_1B_2}(y,b,b')\tr_{A_1A_2} \left[W^{A_1A_2B_1B_2}\left(\sum_x\xi^{A_1A_2}(x,a) \right)\right]\right\}.
\end{split}	
\end{displaymath}
The process observed by Bob is therefore described by the reduced matrix
\begin{equation}
		\label{reduced}
		\overline{W}^{B_1B_2}(a):=\tr_{A_1A_2} \left[W^{A_1A_2B_1B_2}\left(\sum_x\xi^{A_1A_2}(x,a)\right)\right].
\end{equation}
The matrix $\sum_x\xi^{A_1A_2}(x,a)$ represents the CPTP map performed by Alice when the outcomes of her measurement are ignored (the explicit dependence on $a$ accounts for the possibility of signalling). Using \eqref{alice} we find $\sum_x\xi^{A_1A_2}(x,a) = \frac{1}{2} \id^{A_1}\otimes\left[\id+(-1)^a\sigma_z \right]^{A_2}$, which, plugged into Eq. \eqref{reduced} together with Eq. \eqref{quantum}, gives
\begin{equation}
		\label{bobstate}
		\overline{W}^{B_1B_2}(a) = \frac{1}{2}\left[\id + (-1)^a\frac{1}{\sqrt{2}}\sigma_z \right]^{B_1}\otimes\id^{B_2}.
\end{equation}
When this is measured with the map \eqref{bobreceive}, we find

\begin{gather}
	P(y|ab, b'=1)= \tr \left[\eta_1^{B_1B_2}(y,b) \overline{W}^{B_1B_2}(a)\right] =\frac{1}{2}\left[ 1 + \frac{(-1)^{y+a}}{\sqrt{2}}\right],
\end{gather}
from which we obtain $P(y=a|b'=1)=\frac{2+\sqrt{2}}{4}$.

Consider now the case when $b'=0$. When Bob's outcomes are ignored, he performs the CPTP map described by $\sum_y\eta^{B_1B_2}(y,b)=\frac{1}{2} \left[\id+(-1)^b\sigma^{B_1}_x\sigma_z^{B_2} \right]$. From this we can calculate, as in the previous case, the effective state received by Alice, which is
\begin{equation}
		\label{alicestate}
		\overline{W}^{A_1A_2}(b,b'=0) = \frac{1}{2}\left[\id + (-1)^b\frac{1}{\sqrt{2}}\sigma_z \right]^{A_1}\otimes\id^{A_2},
\end{equation}
from which we find $P(x=b|b'=0)=\frac{2+\sqrt{2}}{4}$. In conclusion, the protocol described yields the probability of success \eqref{success}, which proves that the process in Eq.~\eqref{quantum} is not causally separable.

\subsection{{Casual order in the classical limit}}

Let us now show that in the classical limit all correlations are causally ordered. Classical operations can be described by transition matrices $M^{(ki)}_j = P(k,j|i)$, where $P(k,j|i)$ is the conditional probability that the measurement outcome $j$ is observed and the classical output state $k$ is prepared given that the input state is $i$. They can be expressed in the quantum formalism as CP maps diagonal in a fixed (`pointer') basis, and the corresponding CJ matrices are $M_j= \sum_{ki}M^{(ki)}_j |i\ket\bra i|^{A_1}\otimes |k\ket\bra k|^{A_2}$. In order to express arbitrary bipartite probabilities of classical operations, it is sufficient to consider process matrices of the standard form
\begin{equation}
W^{A_1A_2B_1B_2} = \frac{1}{d_{A_1}d_{B_1}}\left(\id + \sigma^{B\npreceq A}+ \sigma^{A\npreceq B}\right),
\label{form}
\end{equation}
where $\sigma^{B\npreceq A}$ and $\sigma^{A\npreceq B}$ are diagonal in the pointer basis. Probabilities are still given by
\begin{equation}
	P\left(\mathcal{M}^A_i, \mathcal{M}^B_j\right) = \tr \left[W^{A_1A_2B_1B_2}\left(M^{A_1A_2}_i\otimes M_j^{B_1B_2}\right)\right].
\end{equation}

We will show that any such diagonal process matrix can be written in the form
\begin{equation}
W^{A_1A_2B_1B_2} = \frac{1}{d_{A_1}d_{B_1}}\left(\rho^{A_1A_2B_1}+ \rho^{A_1B_1B_2}\right),
\label{form2}
\end{equation}
where $\rho^{A_1A_2B_1}$ and $\rho^{A_1B_1B_2}$ are positive semidefinite matrices. This is sufficient to conclude that $W^{A_1A_2B_1B_2}$ is causally separable. Indeed, if $W^{A_1A_2B_1B_2}$ could be written in the form \eqref{form2}, we know that $\rho^{A_1A_2B_1}$ would not contain Hilbert-Schmidt terms of the types $A_1A_2$ or $A_2$ (which are not allowed in a process matrix), since by assumption these terms are not part of $W^{A_1A_2B_1B_2}$. Therefore, the matrix
\begin{gather}
W^{B\npreceq A}\equiv \frac{\rho^{A_1A_2B_1}}{\tr\rho^{A_1A_2B_1}} d_{A_2}d_{B_2},
\end{gather}
which is positive semidefinite, has trace $d_{A_2}d_{B_2}$, and contains only terms of the allowed types, would be a valid process matrix with no signalling from $B$ to $A$. Similarly,
\begin{gather}
W^{A\npreceq B}\equiv \frac{\rho^{A_1B_1B_2}}{\tr\rho^{A_1B_1B_2}} d_{A_2}d_{B_2}
\end{gather}
would be a valid process matrix with no signalling from $A$ to $B$. The whole process matrix could then be written in the causally separable form
\begin{equation}
W^{A_1A_2B_1B_2} = q W^{B\npreceq A} + (1-q) W^{A\npreceq B},
\label{form3}
\end{equation}
where
\begin{gather}
q\equiv\frac{\tr\rho^{A_1A_2B_1}}{d_{A_1}d_{A_2}d_{B_1}d_{B_2}}.
\end{gather}
Note that $0\leq q \leq 1$ since $\rho^{A_1A_2B_1}$ and $\rho^{A_1B_1B_2}$ in Eq.~\eqref{form2} are positive semidefinite and $\tr W^{A_1A_2B_1B_2}=d_{A_2}d_{B_2}$.

To prove Eq.~\eqref{form2}, we will construct $\rho^{A_1A_2B_1}$ and $\rho^{A_1B_1B_2}$ from the general form in Eq.~\eqref{form}. Let the minimum eigenvalue of $\sigma^{B\npreceq A}+ \sigma^{A\npreceq B}$ be $m$. Since $W^{A_1A_2B_1B_2}$ is positive semidefinite and $\sigma^{B\npreceq A}+ \sigma^{A\npreceq B}$ is traceless, we have $m \in [-1,0]$. Define the matrices
\begin{gather}
\kappa^{A_1A_2B_1}=-m\id +\sigma^{B\npreceq A},\\
\kappa^{A_1B_1B_2}=\sigma^{A\npreceq B}.
\end{gather}
The full process matrix can then be written
\begin{gather}
W^{A_1A_2B_1B_2} = \frac{1}{d_{A_1}d_{B_1}}\left((1+m)\id+\kappa^{A_1A_2B_1} +\kappa^{A_1B_1B_2}\right),
\label{form4}
\end{gather}
where $\kappa^{A_1A_2B_1} +\kappa^{A_1B_1B_2}$ is positive semidefinite.

We are now going to modify $\kappa^{A_1A_2B_1}$ and $\kappa^{A_1B_1B_2}$ by adding matrices of the form $\kappa^{A_1B_1}$ to $\kappa^{A_1A_2B_1}$ and subtracting them from $\kappa^{A_1B_1B_2}$ (therefore leaving $\kappa^{A_1A_2B_1} +\kappa^{A_1B_1B_2}$ unchanged), until we transform both $\kappa^{A_1A_2B_1}$ and $\kappa^{A_1B_1B_2}$ in Eq.~\eqref{form4} into positive semidefinite matrices.

Denote the pointer basis of system $X$ by $|i\rangle^X$, $i=1,...,d_X$, $X = A_1, A_2,B_1, B_2$. All matrices we consider are diagonal in the basis $\{|i\rangle^{A_1}|j\rangle^{A_2}|k\rangle^{B_1}|l\rangle^{B_2}\}$. Let $m_1(i,j,k,l)$ denote the eigenvalues of $\kappa^{A_1A_2B_1}$ corresponding to the eigenvectors $|i\rangle^{A_1}|j\rangle^{A_2}|k\rangle^{B_1}|l\rangle^{B_2}$, and let $m_2(i,j,k,l)$ be the eigenvalues of $\kappa^{A_1B_1B_2}$ corresponding to the same vectors. For every $i$ and $k$, we do the following. Define
\begin{gather}
\tilde{m}_1(i,k)=\min_{j,l} m_1(i,j,k,l),\\
\tilde{m}_2(i,k)=\min_{j,l} m_2(i,j,k,l).
\end{gather}
Note that $m_1(i,j,k,l)$ do not depend on $l$ since $\kappa^{A_1A_2B_1}$ acts trivially on $B_2$, and similarly $m_2(i,j,k,l)$ do not depend on $j$. This means that for given $i$ and $k$, the minimum of the eigenvalues of $\kappa^{A_1A_2B_1}+\kappa^{A_1B_1B_2}$ for all eigenvectors of the type $|i\rangle^{A_1}|j\rangle^{A_2}|k\rangle^{B_1}|l\rangle^{B_2}$ is equal to $\tilde{m}_1(i,k)+\tilde{m}_2(i,k)$. But by construction $\kappa^{A_1A_2B_1} +\kappa^{A_1B_1B_2}$ is positive semidefinite, so we have
\begin{gather}
\tilde{m}_1(i,k)+\tilde{m}_2(i,k)\geq 0.
\label{mm}
\end{gather}
Now, if both $\tilde{m}_1(i,k)$ and $\tilde{m}_2(i,k)\}$ are non-negative, we will not modify $\kappa^{A_1A_2B_1}$ and $\kappa^{A_1B_1B_2}$. However, if one of these numbers is negative, say $\tilde{m}_1(i,k)<0$ (both cannot be negative due to \eqref{mm}), we will add the term $-\tilde{m}_1(i,k)|i\rangle\langle i|^{A_1}\otimes \id^{A_2}\otimes |k\rangle\langle k|^{B_1}\otimes \id^{B_2}$ to $\kappa^{A_1A_2B_1}$ and subtract the same term from $\kappa^{A_1B_1B_2}$. After this step, the modified $\kappa^{A_1A_2B_1}$ is such that the eigenvalues $m_1(i,j,k,l)$ have been changed to $m_1(i,j,k,l)-\tilde{m}_1(i,k)\geq \tilde{m}_1(i,k)-\tilde{m}_1(i,k)=0$, i.e. $\kappa^{A_1A_2B_1}$ does not have any more negative eigenvalues $m_1(i,j,k,l)$ for the given $i$ and $k$. The same holds for $\kappa^{A_1B_1B_2}$ since the eigenvalues $m_2(i,j,k,l)$ change to $m_2(i,j,k,l)+\tilde{m}_1(i,k)\geq \tilde{m}_2(i,k)+\tilde{m}_1(i,k)\geq 0$. In other words, the eigenvalues of the modified ${\kappa}^{A_1A_2B_1}$ and ${\kappa}^{A_1B_1B_2}$ satisfy
\begin{gather}
m_1(i,j,k,l), \hspace{0.1cm} m_2(i,j,k,l)\geq 0, \hspace{0.2cm} \forall j,l.
\end{gather}
By performing this procedure for all $i$ and $k$, we eventually transform ${\kappa}^{A_1A_2B_1}$ and ${\kappa}^{A_1B_1B_2}$ into matrices all of whose eigenvalues are non-negative. Denote the resultant positive semidefinite matrices by $\tilde{\kappa}^{A_1A_2B_1}$ and $\tilde{\kappa}^{A_1B_1B_2}$. We can now add the term $(1+m)\id$ in Eq.~\eqref{form4} for instance to $\tilde{\kappa}^{A_1A_2B_1}$ (recall that $m\in[-1,0]$), defining the positive semidefinite matrices
\begin{gather}
 \rho^{A_1A_2B_1}\equiv(1+m)\id+\tilde{\kappa}^{A_1A_2B_1},\\
 \rho^{A_1B_1B_2}\equiv\tilde{\kappa}^{A_1B_1B_2}.
\end{gather}
We thus arrive at the desired form \eqref{form2} which implies \eqref{form3} as argued above.


\begin{thebibliography}{1}

\bibitem{Fivel}
Fivel, D. I. How interference effects in mixtures determine the rules of quantum mechanics. \textit{Phys. Rev. A} \textbf{50}, 2108-2119 (1994).

\bibitem{Zeilinger} Zeilinger, A. A Foundational Principle for Quantum Mechanics. \textit{Found. Phys.} \textbf{29}, 631-643 (1999).

\bibitem{Hardy} Hardy, L. Quantum Theory From Five Reasonable Axioms. Preprint at arXiv:quant-ph/0101012 (2001).

\bibitem{CBH} Clifton, R., Bub, J. \& Halvorson, H. Characterizing Quantum Theory in Terms of Information-Theoretic Constraints. \textit{Found. Phys.} \textbf{33}, 1561-1591 (2003).

\bibitem{goyal}
Goyal, P., Knuth, K. H. \& Skilling, J. Origin of Complex Quantum Amplitudes and Feynman's Rules. \textit{
Phys. Rev. A} \textbf{81}, 022109 (2010).

\bibitem{DB} Dakic, B. \& Brukner, {\v C}. Quantum Theory and Beyond: Is Entanglement Special?
\textit{Deep Beauty: Understanding the Quantum World Through Mathematical Innovation}, Eds. Halvorson, H.  (Cambridge Univ. Press, 2011).

\bibitem{Mas} Masanes, L. \& M\"{u}ller, M. P. A derivation of quantum theory from physical requirements. \textit{New J. Phys.} \textbf{13}, 063001 (2011).

\bibitem{Chiribella2} Chiribella, G., D'Ariano, G. M. \& Perinotti, P. Informational derivation of Quantum Theory.
\textit{Phys. Rev. A} \textbf{84}, 012311 (2011).



\bibitem{hardyqg} Hardy, L. Probability Theories with Dynamic Causal Structure: A New Framework for Quantum Gravity. Preprint at arXiv:gr-qc/0509120 (2005).








\bibitem{Bell} Bell, J. S. On the Einstein Podolsky Rosen Paradox.
\textit{Physics} \textbf{1}, 3, 195-200 (1964).

\bibitem{PR} Popescu, S. \& Rohrlich, D. Quantum nonlocality as an axiom.
\textit{Found. Phys.} \textbf{24}, 379-385 (1994).


\bibitem{BGNP} Beckman, D., Gottesman, D., Nielsen, M. A. \& Preskill, J.
Causal and localizable quantum operations.
\textit{Phys. Rev. A} \textbf{64}, 052309 (2001).

\bibitem{BHK} Barrett, J., Hardy, L. \& Kent, A.
Nonlocal correlations as an information-theoretic resource.
\textit{Phys. Rev. A} \textbf{71}, 022101 (2005).

\bibitem{werner} Arrighi, P. Nesme, V. \& Werner, R.
Unitarity plus causality implies localizability,
\emph{J. Comput. Syst. Sci.} \textbf{77}, 2, 372-378 (2011).

\bibitem{infocaus} Paw\l owski, M. \textit{et al.}
Information causality as a physical principle.
\textit{Nature} \textbf{461}, 1101-1104 (2009).

\bibitem{macrolocality} Navascues, M. \& Wunderlich, H. A glance beyond the quantum model.
\textit{Proc. Roy. Soc. Lond. A} \textbf{466}, 881-890 (2009).

\bibitem{wolf}
Wolf, M. M., Perez-Garcia, D. \& Fernandez, C. Measurements Incompatible in Quantum Theory Cannot Be Measured Jointly in Any Other No-Signaling Theory. \textit{Phys. Rev. Lett.} \textbf{103}, 230402 (2009).

\bibitem{Barnum} Barnum, H., Beigi, S., Boixo, S., Elliott, M. B. \& Wehner, S.
Local Quantum Measurement and No-Signaling Imply Quantum Correlations. \textit{Phys. Rev. Lett.} \textbf{104}, 140401 (2010).

\bibitem{Acin} Acin, A. \textit{et al.}
Unified Framework for Correlations in Terms of Local Quantum Observables. \textit{Phys. Rev. Lett.} \textbf{104}, 140404 (2010).





\bibitem{dewitt}
DeWitt, B. S. Quantum Theory of Gravity. I. The Canonical Theory. \textit{Phys. Rev.} \textbf{160}, 1113-1148 (1967).


\bibitem{perestime}
Peres, A. Measurement of time by quantum clocks. \textit{Am. J. Phys.} \textbf{48}, 552-557 (1980).

\bibitem{wooters}
Wooters, W. K. ``Time'' replaced by quantum correlations. \textit{Int. J. Theor. Phys.} \textbf{23}, 701-711 (1984).

\bibitem{isham}
Isham, C. J. \& Kuchar, K. V. Representations of Space-time Diffeomorphisms. 2. Canonical Geometrodynamics. \textit{
Ann. Phys.} \textbf{164}, 2, 316-333 (1985).

\bibitem{rovelli}
Rovelli, C. Quantum mechanics without time: A model. \textit{Phys. Rev. D} \textbf{42}, 2638-2646 (1990).

\bibitem{gambini}
Gambini, R., Porto, R. A. \& Pullin, J. A relational solution to the problem of time in quantum mechanics and quantum gravity: a fundamental mechanism for quantum decoherence. \textit{New J. Phys.} \textbf{6}, 45 (2004).














\bibitem{instrument}
Davies, E. \& Lewis, J.
An operational approach to quantum probability.
\textit{Comm. Math. Phys.} \textbf{17}, 239-260 (1970).





\bibitem{nielsen_chuang}
Nielsen, M. A. \& Chuang, I. L.
\textit{Quantum computation and quantum information}, (Cambridge University Press, Cambridge, 2000).



\bibitem{jam} Jamio{\l}kowski, A. Linear transformations which preserve trace and positive semidefiniteness of operators. \textit{Rep. Math. Phys.} \textbf{3}, 4, 275-278 (1972).

\bibitem{choi} Choi, M.-D. Completely positive linear maps on complex matrices. \textit{
Lin. Alg. Appl.} \textbf{10}, 285-290 (1975).



\bibitem{networks}
Chiribella, G., D�Ariano,  G. M. \& Perinotti, P. Theoretical framework for quantum networks.
\textit{Phys. Rev. A} \textbf{80}, 022339 (2009).






\bibitem{CHSH} Clauser, J. F., Horne, M. A., Shimony, A. \& Holt, R. A. Proposed experiment to test local hidden-variable theories. \textit{Phys. Rev. Lett.} \textbf{23}, 880-884 (1969).




\bibitem{Finkelstein} Finkelstein, D. Space-time code. \textit{
Phys. Rev.} \textbf{184}, 1261-1271 (1968).

\bibitem{oriti} Oriti, D. \textit{Approaches to Quantum Gravity: Toward a New Understanding of Space, Time and Matter}, (Cambridge Univ. Press, Cambridge, 2009).

\bibitem{glimmers}  Piazza, F. Glimmers of a Pre-geometric Perspective. \textit{Found. Phys.} \textbf{40}, 239-266 (2010).


\bibitem{Zurek} Zurek, W. H. Decoherence and the transition from quantum to classical. \textit{Phys. Today} \textbf{44}, 36-44 (1991).

\bibitem{Kofl2007}
Kofler, J. \& Brukner, {\v C}. Classical world arising out of quantum physics under the restriction of coarse-grained measurements. \textit{Phys. Rev. Lett.} \textbf{99}, 180403 (2007).


\bibitem{Bomb87} Bombelli, L.,
Lee, J. H., Meyer, D. \& Sorkin, R. Space-time as a causal set. \textit{
Phys. Rev. Lett.} \textbf{59}, 521-524 (1987).

\bibitem{dari} D'Ariano, G. M. \& Tosini, A. Space-time and special relativity from causal networks. Preprint at  arXiv:1008.4805 (2010).

\bibitem{Hawking} Hawking, S. W., King, A. R. \& McCarthy, P. J. A new topology for curved space-time which incorporates the causal, differential, and conformal structures. \textit{J. Math. Phys.} \textbf{17}, 174-181 (1976).

\bibitem{Malament} Malament, D. B. The class of continuous timelike curves determines the topology of spacetime. \textit{J. Math. Phys.} \textbf{18}, 1399-1404 (1977).



\bibitem{Goedel} G\"{o}del, K.  An Example of a New Type of Cosmological Solution of Einstein's Field Equations of Gravitation. \textit{Rev. Mod. Phys.} \textbf{21}, 447-450 (1949).


\bibitem{Deutsch} Deutsch, D. Quantum mechanics near closed timelike lines. \textit{Phys. Rev. D} \textbf{44}, 3197-3217 (1991).

\bibitem{Bennett2}
Bennett, C. H. \textit{Talk at QUPON, Vienna, May 2005} (based on an unpublished work with Schumacher, B.). Presentation at $\langle$http://www.research.ibm.com/people/b/bennetc$\rangle$ (2005).

\bibitem{greenberger}
Greenberger, D. M. \& Svozil, K. Quantum Theory Looks at Time Travel.
\textit{Quo Vadis Quantum Mechanics?}, Eds. Elitzur, A., Dolev, S. \& Kolenda, N. (Springer Verlag, Berlin, 2005).

\bibitem{Lloyd} Lloyd, S. \textit{et al.}
 Closed timelike curves via post-selection: theory and experimental demonstration. \textit{Phys. Rev. Lett.} \textbf{106}, 040403 (2011).


\bibitem{Bennett1}
Bennett, C. H., Leung, D., Smith, G., Smolin, J. A.
Can Closed Timelike Curves or Nonlinear Quantum Mechanics Improve Quantum State Discrimination or Help Solve Hard Problems? \textit{Phys. Rev. Lett.} \textbf{103}, 170502 (2009).


\bibitem{CTC} Brun, T. A. \& Wilde, M. M. Perfect state distinguishability and computational speedups with postselected closed timelike curves. \textit{Found. Phys.} \textbf{42}, 3, 341-361 (2012).


\bibitem{Novikov} Friedman, J. \textit{et al.}
Cauchy problem in spacetimes with closed timelike curves. \textit{Phys. Rev. D} \textbf{42}, 1915-1930 (1990).











\bibitem{chiribella3} 
Chiribella, G., D'Ariano, G. M., Perinotti, P. \& Valiron, B. Beyond causally ordered quantum computers. Preprint at
arXiv:0912.0195 (2009).


\bibitem{popt} Barnum, H., Fuchs, C. A., Renes, J. M. \& Wilce, A. Influence-free states on compound quantum systems.
Preprint at arXiv:quant-ph/0507108 (2005).



\bibitem{Gleason} Gleason, A. M. Measures on the closed subspaces of a Hilbert space. \textit{J. Math. Mech.} \textbf{6}, 885-893 (1957).

\bibitem{CFMR} Caves, C. M., Fuchs, C. A., Manne, K. K. \& Renes, J. M. Gleason-Type Derivations of the Quantum Probability Rule for Generalized Measurements. \textit{Found. Phys.} \textbf{34}, 2, 193-209 (2004).



\end{thebibliography}
\end{document}